\documentclass[journal=jpcm,10pt]{iopart}
\usepackage{graphicx,epsfig}
\usepackage{amsfonts}
\usepackage{iopams}
\usepackage{cite}
\usepackage[nointegrals]{wasysym} 
\usepackage[normalem]{ulem}
\usepackage{bm}
\usepackage{cuted}
\usepackage{amssymb,xcolor}
\usepackage{graphicx}
\usepackage[colorlinks, allcolors=blue]{hyperref}
\usepackage{array,multirow}
\newcolumntype{C}[1]{>{\centering\arraybackslash}m{#1}}
\begin{document}
\title{High Temperature Superconductivity in the Cuprates:
Phenomena from a Theorist$^\prime$s Point of View}
\author{T V Ramakrishnan}
\address{Indian Institute of Science, Bangalore
 and JNCASR, Jakkur, Bangalore}

\begin{abstract}
I present a selection of experimental results on metallic cuprates, both above the superconducting 
 transition temperature $T_c$ (often called the strange metal state) and in the superconducting state. It highlights this still poorly understood part of the physical world. After an introduction, I talk briefly about the pseudogap regime and about the unusual linear resistivity phenomenon. Several empirical correlations between observed quantities are mentioned, e.g. $T_c$ and superfluid density (Uemura), $T_c$ and next nearest neighbour hopping, slope of the linear resistivity and $T_c$. In the belief that a comprehensive explanation may need an understanding of the extremely strongly correlated metal, a few initial steps in this direction are outlined.
\end{abstract}

\maketitle
\ioptwocol

\section{Introduction}
\label{sec:introduction}
It is a great honour to participate in a meeting marking a hundred years of the revolutionary work 
 of Satyendranath Bose. Bose$^\prime$s pioneering work on an ideal gas of photons showed us how to count states of a collection of identical, free, massless, spinful, quantum particles. It marked a turning point; quantum statistical physics was born. For this reason, the great theoretical physicist Landau regarded him as one of the founders of quantum physics along with Bohr, Heisenberg, Schroedinger and Dirac.\footnote{Landau had a logarithmic ‘genius scale’ for physicists, ranging from 0 to 5. Newton was at 0 on this scale and Einstein was at 0.5. Bose, along with foundational figures of quantum physics like Bohr, Schrodinger, Heisenberg, Dirac… was at 1. Landau put himself at 2.5, and was finally promoted to 2.}

As is well known, Einstein applied the quantum statistics of Bose \cite{Bose1924} to particles with nonzero 
 rest mass. He showed that below a certain temperature, inevitably, a macroscopic fraction of them condenses 
 into a single quantum state \cite{Einstein1924}; this is Bose (or Bose Einstein) condensation. Very soon after the discovery of superfluidity in Helium (1937), London identified superfluidity with Bose condensation \cite{London1938}. Later, he argued that superconductivity is due to macroscopic quantum coherence, a phenomenon akin to Bose condensation \cite{London1948,London1990}. On the face of it, macroscopic quantum coherence in a metal appears unlikely; the constituent electrons are subject to the Pauli exclusion principle so that no two of them can be in a single quantum state, let alone a very large number being in the same state. Their binding into pairs (Cooper pairs) which are Bose like objects and the description of a superconductor as 
 a phase coherent collection of these Cooper pairs is the celebrated Bardeen, Cooper and Schrieffer (BCS) 
 theory of superconductivity \cite{Bardeen1957}, proposed in 1957. It seems even more improbable that cuprates could be superconducting,
since additionally, electrons in them repel each other strongly and
locally (strong electron correlation, indicated by large
positive $U$ in the Hubbard model) \cite{Sheshadri2023}. However, very soon
after the discovery of high temperature superconductivity
in the cuprates \cite{Bednorz1986} in 1986, Anderson \cite{Anderson1987} argued that
superconductivity can occur in purely and strongly
repulsive electron systems, and proposed a mechanism
(resonating valence bond or RVB) for it.

Two unexpected discoveries in the physics of quantum
matter dating to the $1980$s, namely the quantum Hall
effect \cite{Klitzing1980} and high temperature superconductivity in the
cuprates \cite{Bednorz1986} have given birth to new directions in condensed
matter physics. The recognition of topology as a widely
prevalent and robust feature of condensed matter
systems with unexpected consequences flowed from the
former. The basic electronic nature of the latter, namely of
the cuprates and their superconductivity, is still a relatively unsettled question while it is possible that the answer has
the potential to open up a large domain of quantum matter
for deeper exploration. Although cuprate phenomena
(both in the nonsuperconducting or 'normal' metallic as
well as in superconducting phases) no longer occupy center
stage in the field of condensed matter physics, a huge
mountain of work has accumulated (there are apparently
more than $2,00,000$ papers on the subject) and surprising
major effects continue to be discovered. There has been
great progress both theoretically and experimentally in the
field. There are many demanding and highly sophisticated
theoretical approaches. However, the broad view seems
to be that we do not quite understand the unusual goings
on in them as a whole.

In this article, I introduce the cuprate family in section \ref{sec-IIA}
and describe some experimental features (Sections \ref {sec-III} to \ref{sec-V})
limited by my knowledge, memory and understanding.
Fortunately, there is a large review literature on the subject; synoptic reviews \cite{Keimer2015}, detailed surveys of parts of
the field e.g. ref. \cite{Vishik2018} and books (e.g. the one edited by Schrieffer
and Brooks) \cite{Robert2007}. Some of the data presented here are more
recent. The correlations presented here are known, but
are put together here in one place for the first time. They
seem to call for a comprehensive and detailed theory in
the strong correlation genre. High temperature
superconductivity in the cuprates probably provides us
with a glimpse of a Bose condensation like phenomenon
which may require a basic departure in quantum physics. \\


\section{Cuprates: the basics\label{sec-IIA}}

Cuprates in which $Cu$ occurs in the electronic configuration
$Cu^{++}$\footnote{namely with a partially filled $d$ shell having $9$ rather
than $10$ $d$ electrons, therefore necessarily with $d$ electron
states near the Fermi energy.} are a large family with more
than $30$ chemically distinct members. To model their
electronic properties, we start with the ‘mother compound’
$La_2CuO_4$. $La_2CuO_4$ in which a fraction of the trivalent $La$
atoms are replaced by divalent $Ba$ was the first discovered
high $T_c$ superconductor\cite{Bednorz1986}; in general high $T_c$
superconductivity in this family results for systems for the
doped systems $La_{2-x}Ae_xCuO_4$ where $x$ is therefore the
amount of hole doping ($\delta$ doping in the semiconductor
language) and $Ae$ is an alkaline earth. The most commonly used alkaline earth is $Sr$.

\begin{figure}[ht!]
\includegraphics[width=\linewidth]{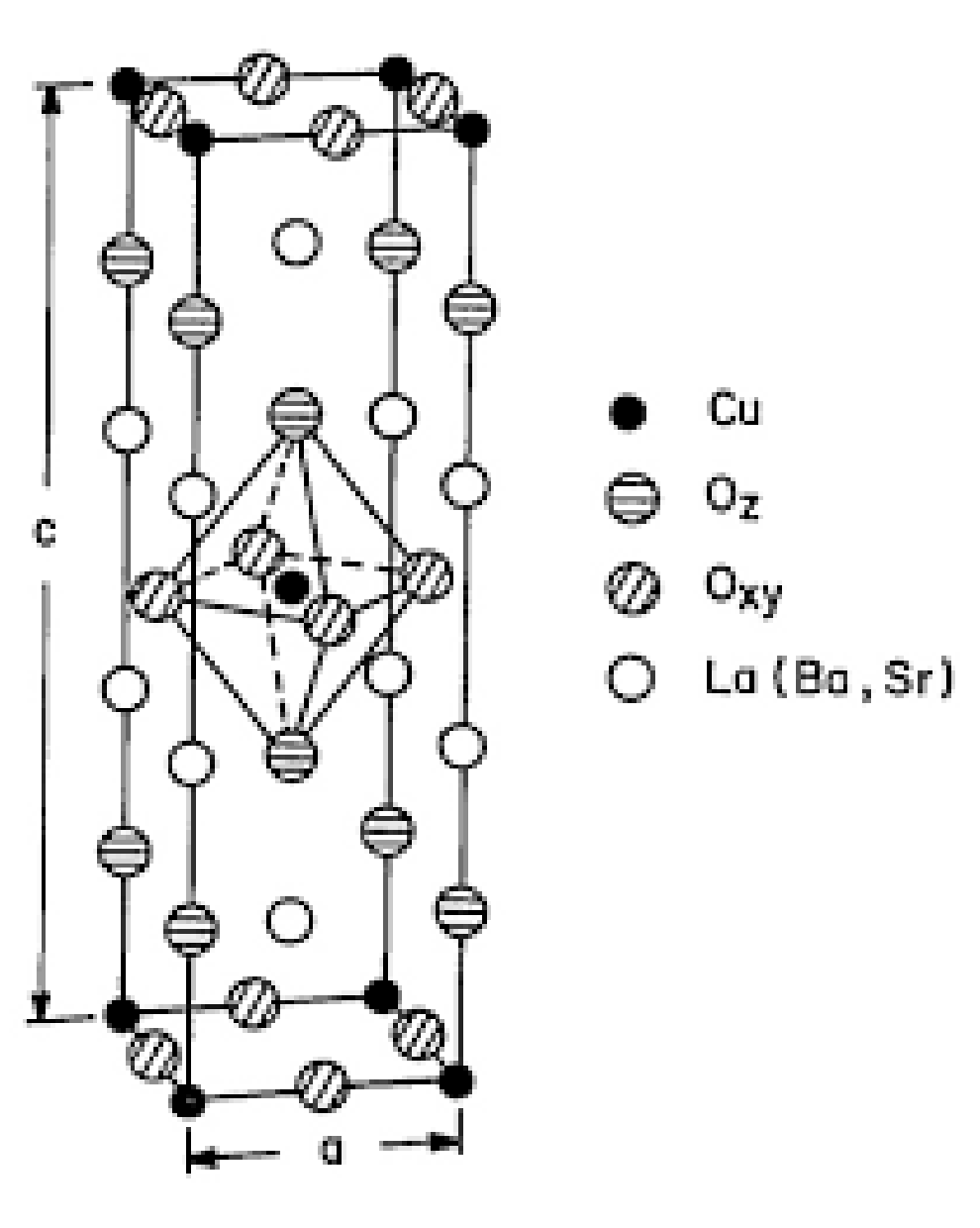}
\caption{Schematic picture of a unit cell of stoichiometric $La_2CuO_4$
crystal.
}
\label{fig1}
\end{figure}

The crystal structure of $La_2CuO_4$ is shown
in Fig. \ref{fig1}. This can be thought of as an ‘enhanced’ perovskite; namely as $ABO_3$ (perovskite) + $AO$ where $A \rightarrow La$ and $B \rightarrow Cu$. (This
is one reason why the solid state chemists Ganguly and
Rao were interested \cite{Ganguly1984} in it around $1984$, and showed that it is an antiferromagnetic Mott insulator). The perovskite is
highly distorted; it is a ‘pointy’ octahedron with $BO$ along
the $c$ axis or $z$ direction being $\approx 2.46A$, and $BO$ in $ab$ or $xy$
plane being $\approx 1.91A$. As a consequence, the system is accurately and most simply thought of as planar layers of
($Cu-O_2$) (square plane with $Cu$ at the corners of the
square, and $O$ atoms at the centre of the shortest $Cu-Cu$
line, namely the square edge) interspersed with $La-O$
layers.
\begin{figure}[ht!]
\includegraphics[width=\linewidth]{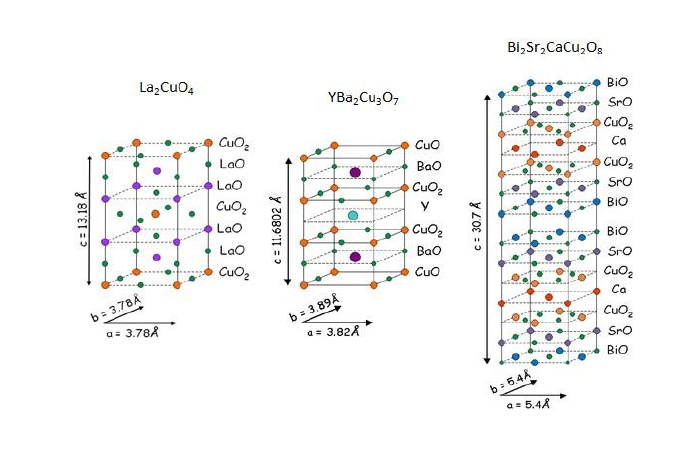}
\caption{Unit cells of three common members of the cuprate family,
namely $La_2CuO_4$, $YBa_2Cu_3O_7$ and $Bi_2Sr_2CaCu_2O_8$.
}
\label{fig2}
\end{figure}

This general feature is apparent from the
schematic unit cell pictures of three well known cuprate
types shown in Fig. \ref{fig2},
namely $La_2CuO_4$ (often called $LCO$; the compound doped with $Sr$ is called $LSCO$), $YBa_{2}Cu_{3}O_{7}$  (called $YBCO$) and
$Bi_2Sr_2CaCu_2O_8$ (called $Bi-2212$). We see well separated layers of $(Cu-O_2)$ ( e.g. the top and bottom and middle layers for $LCO$ in Figs. \ref{fig1} and \ref{fig2}; the bilayers in $YBCO$ and $Bi-2212$ in Fig. \ref{fig2}). The hole doping in these is accomplished by substituting trivalent $La$ with divalent $Sr$ for example, and by oxygen deficiency $(YBCO)$ or excess $(Bi-2212)$. In the latter two systems, greater oxygen deficiency corresponds to lower hole doping $(YBCO)$ and larger oxygen excess to higher doping $(Bi-2212)$. The doping level $x$ is generally estimated using the Presland \cite{Presland1991} formula.

\begin{figure}[ht!]
\includegraphics[width=\linewidth]{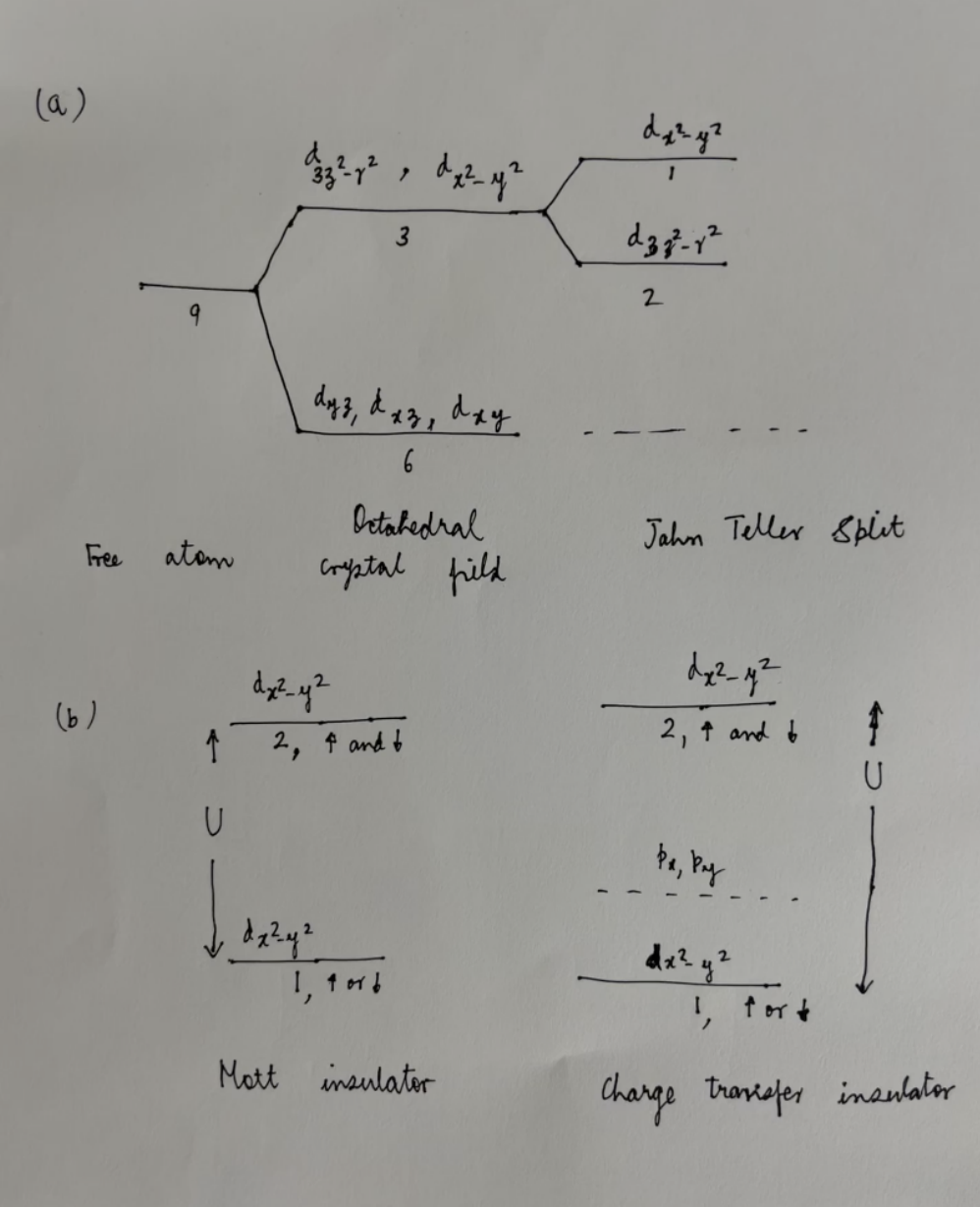}
\caption{(a) $d$ electron levels of $Cu^{++}$ in a free atom, in a perfect octahedral environment, and in an octahedral environment with Jahn Teller splitting of the kind that occurs in the cuprates. Electron
occupation is also indicated. (b) Mott and charge transfer insulators,
shown schematically.
}
\label{fig3}
\end{figure}

Fig. \ref{fig3} (a) depicts the single site $d$ orbital energy level for a free atom (tenfold degenerate), and for an atom in a perfect octahedral symmetry crystal field (which splits this into lower lying degenerate $d_{yz}$, $d_{xz}$, $d_{xy}$ symmetry orbitals accommodating six electrons in all  and higher lying $d_{3z^2-r^2}$,
$d_{x^2-y^2}$ orbitals with a maximum of four electrons in them). It also shows that the Jahn Teller splitting of the perfect octahedral coordination (resulting in a ‘pointy’ octahedron) splits the degeneracy of the upper levels, with the ‘antibonding’ $d_{x^2-y^2}$ symmetry level being higher in energy.
For nine electrons in the $d$ shell, consequently there is one
electron (or hole) in the (local) $d_{x^2-y^2}$ symmetry orbital.
These are the electron and the state we focus on. Fig. \ref{fig3} (b) shows schematically that (as in a stoichiometric cuprate) if there is exactly one $d$ electron per site, electron transport requires that there will necessarily be two electrons on the site to which the
electron hops. This costs an extra energy $U$. If this is large
(larger than the kinetic energy gained by motion), such a
real hopping does not take place, each electron stays put,
and the system is an insulator, a Mott insulator. If there are
unoccupied $p_x$ or $p_y$ states within the Mott-Hubbard gap of energy $ \approx U$ (as in the cuprates) the lowest energy
excitation involves transferring an electron from a $d_{x^2-y^2}$ orbital to
a $p$ orbital and the system is a charge transfer insulator \cite{Zaanen1985}.

With this background, I model the
interesting low electron excitation energy states of the
cuprates by considering only the uncoupled parallel planes
with side sharing square unit cells each with $Cu$ atoms at
corners and $O$ atoms at centers of sides. There are $d_{x^2-y^2}$
local symmetry orbitals situated at $Cu$ sites, $p_x$ and $p_y$
orbitals at $O$ sites. Each orbital can accommodate two
electrons. There is onsite electron repulsion between
electrons in these orbitals; and one has intersite hopping .
A very common description of hole doped cuprates is the
following. The hole migrates to the $p_x$ , $p_y$ orbitals of the plane because of the strong coulomb attraction to the ($Cu-O_2$) collection. Zhang and Rice \cite{Zhang1988} showed (see also
ref. \cite{Ogata2008}) that it hybridizes with the local orbital of $d_{x^2-y^2}$
symmetry and forms a strongly bound spin singlet (or
spinless) hole state.
Therefore, the simplest picture which has a chance of
being realistic is an effective one band Hubbard model
(two $d$ square lattice, $d_{x^2-y^2}$ local symmetry orbital at site $i$) with such holes. This is described by the following Hamiltonian $(H-\mu N_e)$)in a grand canonical ensemble:

\begin{eqnarray}
\label{ssm_eq1}
{H-\mu N_e}=\sum_{i,\sigma} (\varepsilon_d-\mu) a^\dagger_{i \sigma} a_{i \sigma} \sum_{i,j}t_{ij} a^\dagger_{i \sigma} a_{j \sigma} +  \nonumber \\
& & \hspace*{-80pt} U \sum_{i} {n_{i\uparrow}n_{i\downarrow}}
\end{eqnarray}

All the other electronic states have been ‘integrated out’,
and their effect appears in the parameters of this effective
one band Hubbard model. It is perhaps the most
commonly used model for describing the cuprates. Here, $\varepsilon_d$
is the energy of the $d_{x^2-y^2}$ orbital at site $i$, the chemical
potential $\mu$ is such that one has $x$ holes or $(1-x)$ electrons
per site, \footnote{In the literature, $x$ is often referred to as
$p$, the symbol for a hole in semicoductor physics} $t_{ij}$ is the
hopping amplitude from site $i$ to site $j$ and $U$ is the on site Mott-Hubbard repulsion. Very many experimental and theoretical efforts have provided estimates for the parameters of the Hamiltonian. A common one is : $t$ (the nearest neighbour hopping amplitude) $\approx 0.4 eV$, $t^\prime$ (the next nearest neighbour hopping amplitude) $\approx-0.3t$ and $U \approx 4-7 eV$ (see e.g. ref \cite{Sheshadri2023} for a survey as well as a calculation of $U$).

An obvious question is: do we have a theory of
superconductivity in the cuprates in this model? It is a
measure of the fractious nature of the subject that there is
no convergence on this issue. As mentioned above, there
is an RVB mechanism of superconductivity proposed by
Anderson \cite{Anderson1987} and developed by him and Baskaran \cite{Baskaran1987} as well
as a ‘plain vanilla’ strong correlation theory described by
him and a number of collaborators \cite{Anderson2004}. On the other hand,
we know a large number of specific properties of cuprate
superconductors and of the strange metal from which the
superconductivity arises; these have not yet found an
explanation in the theory. In the next few sections
(Sections \ref{sec-III} - \ref{sec-V}) we detail experimental properties whose
comprehensive and coherent explanation would constitute
a complete theory.

\begin{figure}[ht!]
\includegraphics[width=\linewidth]{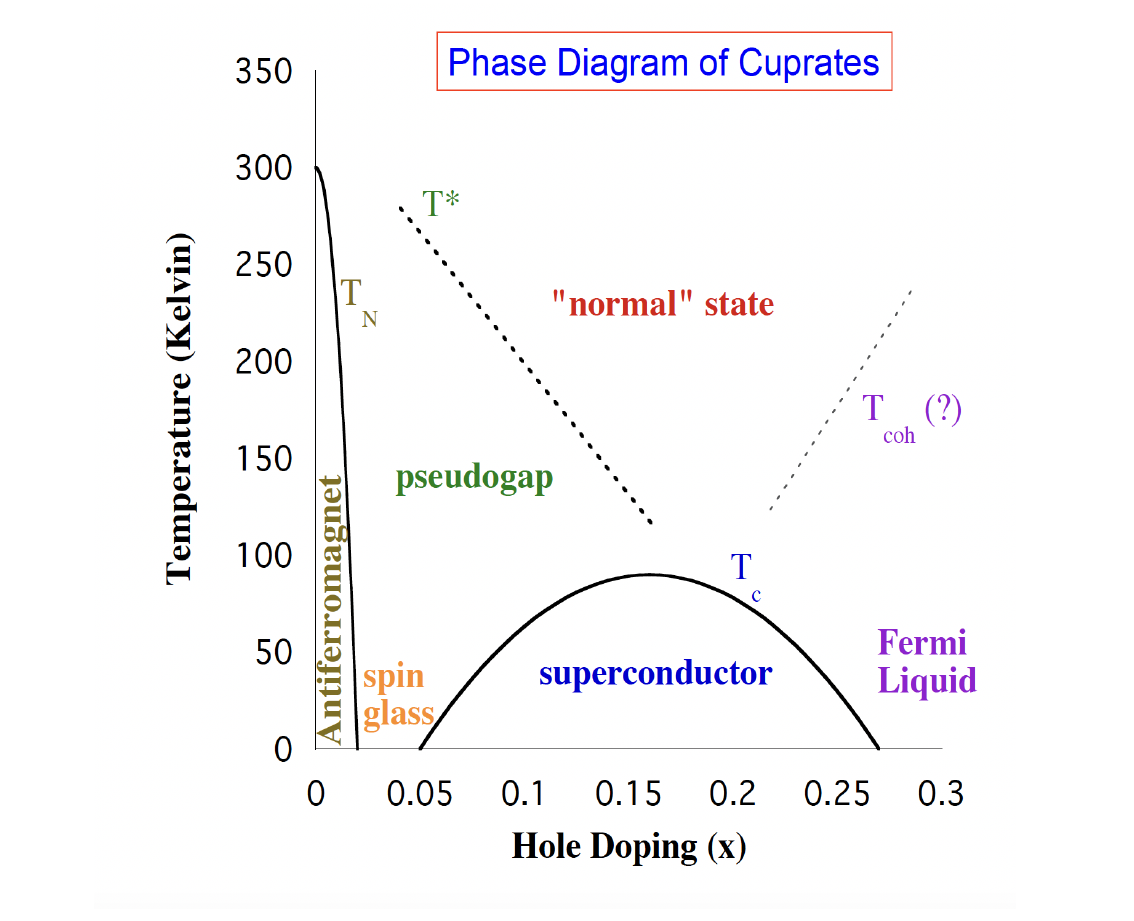}
\caption{Phase diagram of the hole doped cuprates in the hole $x$ (or $p$)
and temperature $T$ plane (from Norman, \cite{Norman2011}).
}
\label{fig4}
\end{figure}

We situate the description of experimental phenomena in the observed phase diagram of hole doped cuprate superconductors. An early,
simplified, universal phase diagram in the hole density $x$
temperature $T$ plane , due to Norman \cite{Norman2011} is shown in Fig. \ref{fig4}. (The hole concentration $x$ in
cuprates is mostly inferred from the highly successful
empirical Presland \cite{Presland1991} formula). The stoichiometric cuprate
$(x=0)$ is a N\'eel antiferromagnetic insulator below $T_N$ . The material is
a paramagnetic insulator above $T_N$ , a signature of the Mott
or correlated insulator. The N\'eel order and the insulating
phase disappear very rapidly with doping; typically, for
$x >0.04$ there is no N\'eel order. There is a very narrow regime of glassy insulating
phase for larger $x$ at low temperatures around this $x$. The
superconducting phase begins at around $x \approx 0.04$ (say $x_{min}$ )
and continues till $x\approx 0.27$ (say $x_{max}$) in a roughly parabolic
dome shaped curve in the $(x,T)$ plane. $T_c$ is maximum for
about $x \approx 0.16$; this is generally called optimum doping $x_{opt}$.
The system is called underdoped for $x<x_{opt}$ and overdoped
for $x>x_{opt}$. Most of the early work concentrated on the
underdoped regime. Recently, there is much more activity
in the overdoped region. Beyond the doping $x=x_{max}$, the
low temperature region is marked Fermi liquid, which is
characterized for example by the imaginary part of the self
energy $\Sigma(\omega,T)$ having a specific quadratic dependence on
$\omega$ and on $T$. This Fermi liquid behaviour continues with increasing temperature and hole density till a dotted curve region marked $T_{coh}$ with a question mark(?),
above which the Fermi liquid becomes incoherent. For
values of $x$ less than $x_{opt}$, we have, above $T_c$ , a region marked pseudogap below a temperature $T^*(x)$ with distinct
properties some of which are mentioned below. Beyond
the green dotted line where the pseudogap ends, there is a
region marked ‘normal’ state. This roughly fan shaped
region is seen to encompass the slightly underdoped as
well as overdoped regimes of $x$. Their properties have
been explored very actively in the last decade or so. The
thermodynamic phases are two, namely the
antiferromagnetically ordered one and the superconductor.
Their boundaries are indicated by full lines. The other
‘phases’ are regions with different distinct characteristic
behaviour crossing over smoothly. There is a controversy
about whether the pseudogap is a distinct thermodynamic
phase ending in a line of critical points separating the
‘normal’ state, or whether it is a crossover regime (for example
whether in the figure, $T^*$ should be indicated by a dotted line or a full
line).
\begin{figure}[ht!]
\includegraphics[width=\linewidth]{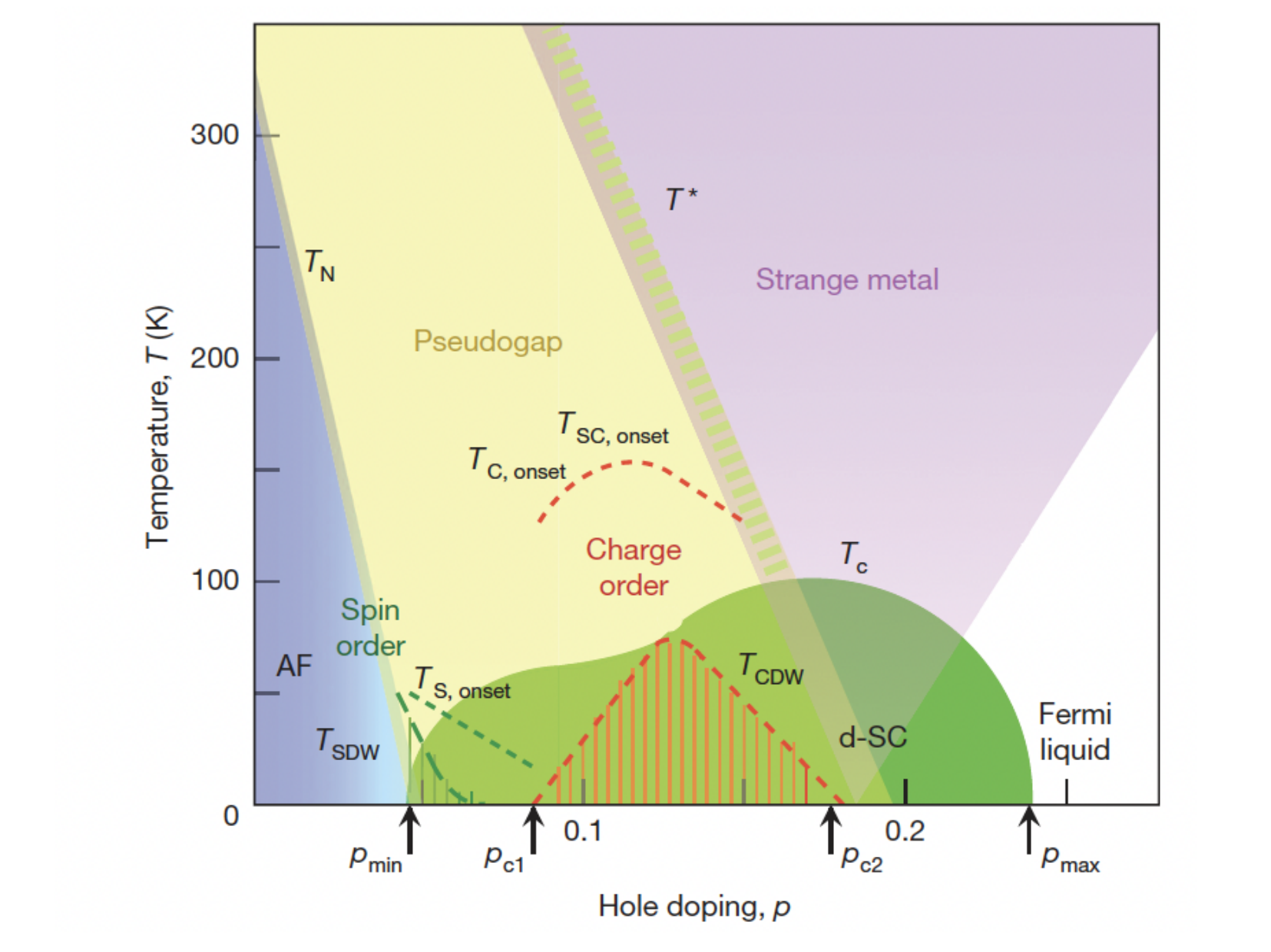}
\caption{Phase diagram of the hole doped cuprates in the hole $x$ (or $p$)
and temperature $T$ plane (from Keimer et al., \cite{Keimer2015}).)
}
\label{fig5}
\end{figure}

A somewhat more realistic and recent phase diagram \cite{Keimer2015} is
shown in Fig. \ref{fig5}. The shape
of the superconducting dome is close to what is actually
seen in $La_{2-x}Sr_xCuO_4$. Many ordering tendencies, specially
in the underdoped region, are indicated, e.g. spin order,
charge (CDW) order. The ‘normal’ state is labelled
as strange metal, which is the present nomenclature for it.

\begin{figure}[ht!]
\includegraphics[width=\linewidth]{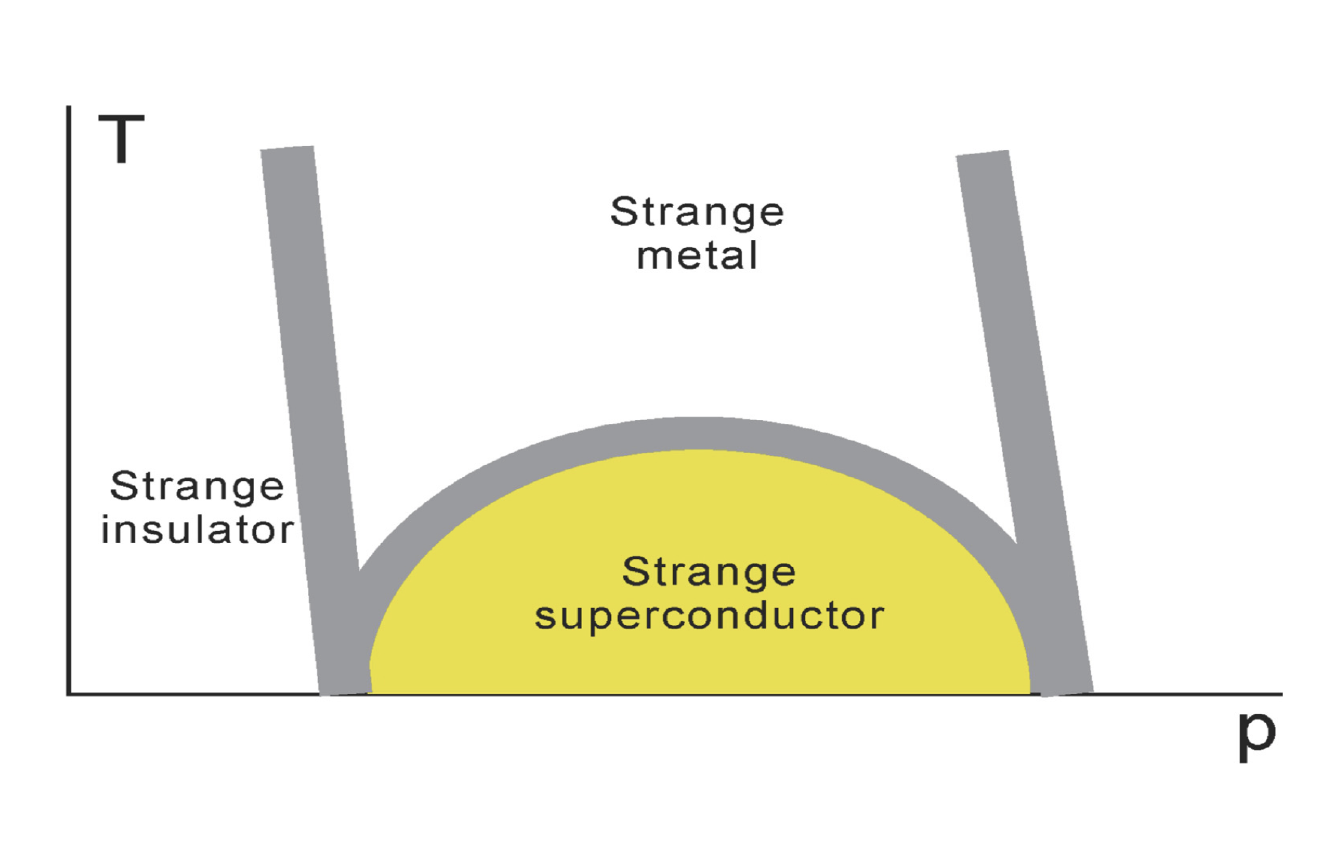}
\caption{A caricature of the hole doped cuprate phase diagram,
emphasizing its strangeness (from Bozovic, \cite{Bovzovic2019}).}
\label{fig6}
\end{figure}
In Fig. \ref{fig6}, I show a schematic version 
of the phase diagram, from an article \cite{Bovzovic2019} by I. Bozovic, an experimentalist who has done experiments on carefully
MBE deposited LSCO samples from underdoped to
overdoped with well defined $x$ (or $p$), for decades. He
argues here that all the three broad regimes in the 
hole doped cuprate phase diagram, namely the insulator, the
metal, and the superconductor are strange. I now describe
some phenomena in two regions, namely the metal and
the superconductor; we see that the properties are indeed
unlike those of ‘conventional’ metals and superconductors.
I focus on experimental results which can be obtained from
the data without too much ‘processing, e.g. resistivity $\rho$
and critical temperature $T_c$ . I also emphasize correlations
between two measured quantities. \\


\section{The Pseudogap\label{sec-III}}
The defining feature of this regime (which is in the
underdoped hole density $x$ regime above $T_c$ and is
bounded by the temperature $T^*(x))$ is the pseudogap
observed in the single particle density of states. 

\begin{figure}[ht!]
\includegraphics[width=\linewidth]{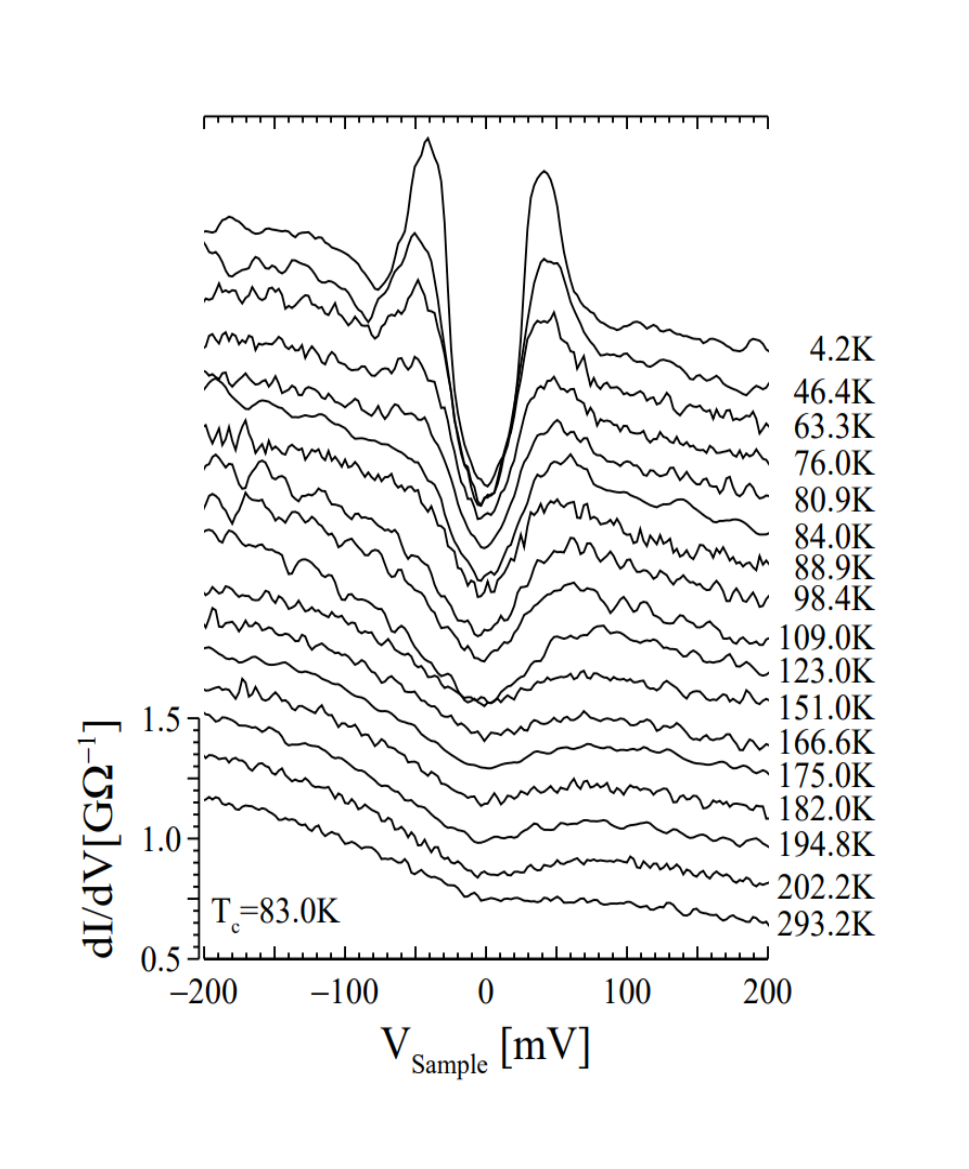}
\caption{Pseudogap in the electron density of states (from Kordyuk,
ref. \cite{Kordyuk2015}).}
\label{fig7}
\end{figure}

Fig. \ref{fig7} (from
a $2015$ review by Kordyuk, ref. \cite{Kordyuk2015}) shows ARPES data for a sample of $Bi-2212$
at different temperatures; the $T_c$ is $83.0K$. We notice a $U$
shaped dip in the inferred density of states (DOS). It is
centred around the Fermi energy and is most prominent at
the lowest measurement temperature $(4.2K)$ where it is
almost a gap. (One can also see a clear particle hole
asymmetry). As temperature increases, the dip becomes
shallower. It does not vanish at $T_c$, where superconductivity
disappears (as happens for the measured DOS of BCS
superconductors, which show such a dip. The DOS there is fitted
well with the BCS theory, and the inferred gap $\Delta(T)$
vanishes at $T_c$). It continues to temperatures much higher
than $T_c$ ; here one can see such a feature till about $166 K$,
which is close to $T^*$ at this doping. Another peculiar
observed feature is that the peak to peak energy (akin to 
$2\Delta(T)$) is roughly the same over this range of temperatures.
The presence of $T^*$ and of the
pseudogap is confirmed by a very large number of
measurements (see ref. \cite{Mueller2017} for a relatively recent review)
such as ARPES, single particle tunnelling, NMR (
magnetic susceptibility), resistivity, neutron scattering,
ultrasound propagation, Kerr effect…. . The range of
values of $T^*$ and the general trend as a function of $x$ are
the same; the actual numbers can and do vary. 
\begin{figure}[ht!]
\includegraphics[width=\linewidth]{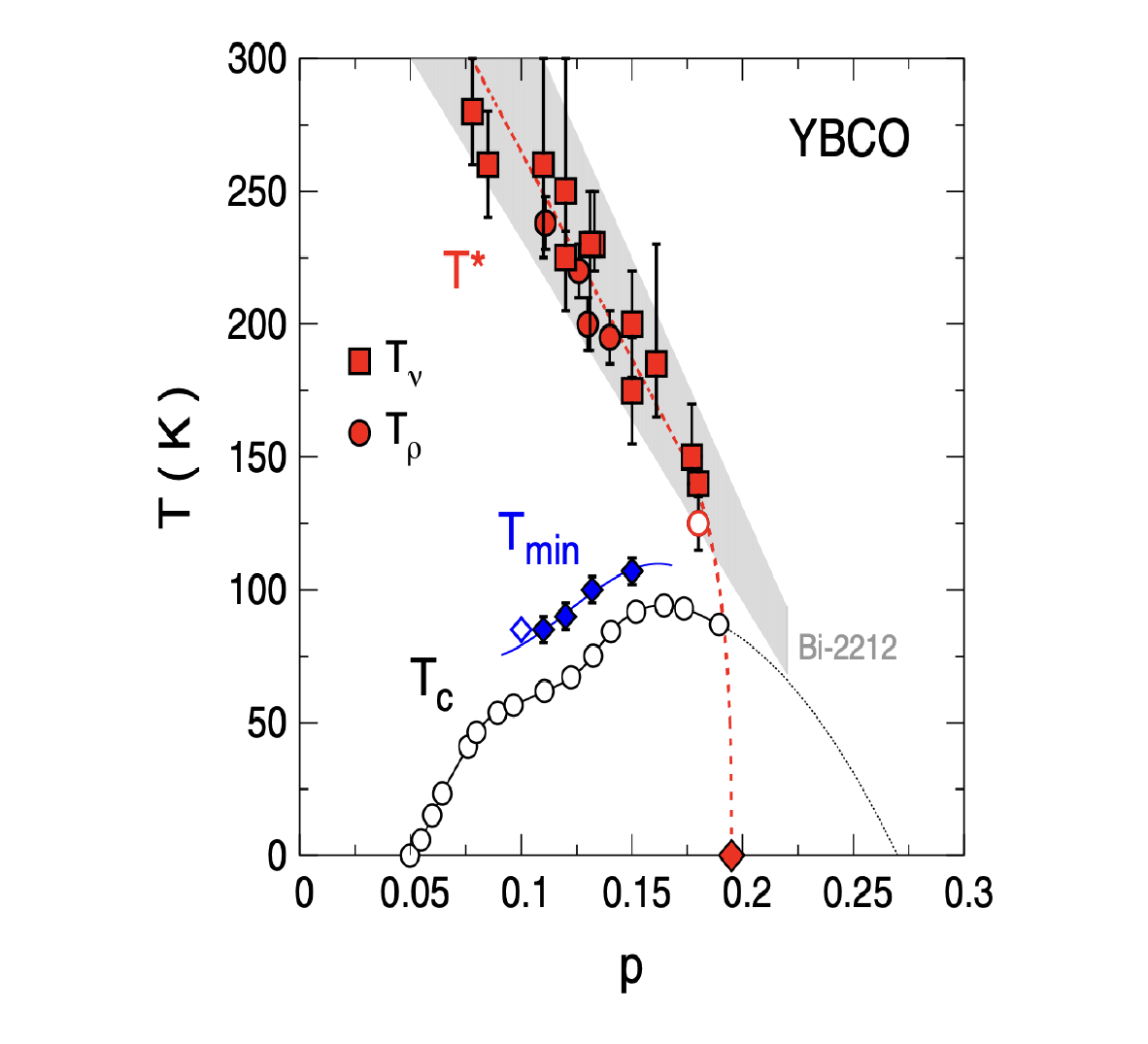}
\caption{The pseudogap temperature $T^*(x)$ as a crossover, from
transport measurements on hole doped YBCO (from Cyr-Choiniere et
al., ref. \cite{Cyr2018}).}
\label{fig8}
\end{figure}

\begin{figure}[ht!]
\includegraphics[width=\linewidth]{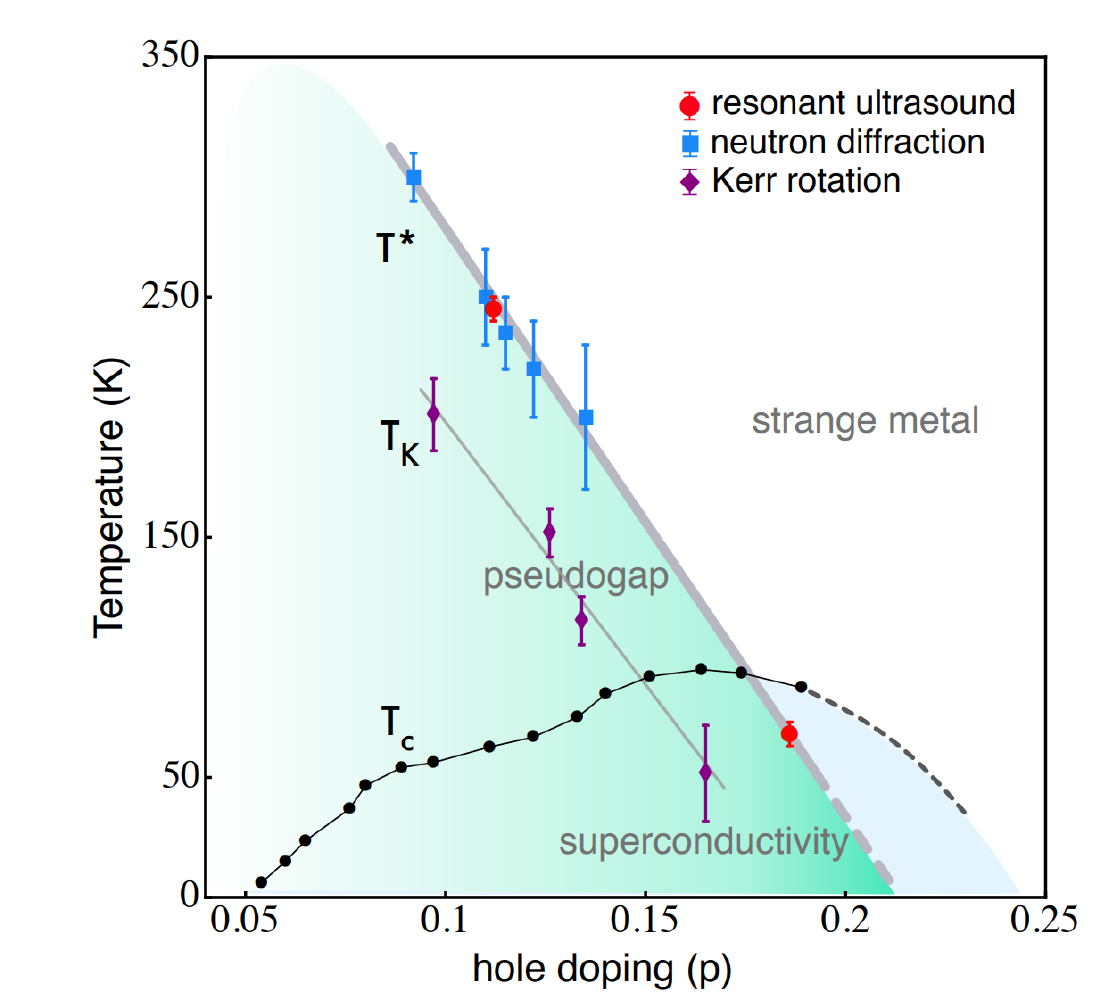}
\caption{The pseudogap temperature $T^*(x)$ as line of phase transitions; hole doped $YBCO$ (from Shekhter et al., \cite{Shekhter2013}).}
\label{fig9}
\end{figure}

Fig. \ref{fig8}
from an article by O. Cyr-
Choiniere et al. \cite{Cyr2018} is an example. The $T^*$ line marking the end of the pseudogap is inferred here for hole doped YBCO from features in the
Nernst effect and in the resistivity.
There is a contrary view that $T^*$ marks a line of phase
transitions rather than a crossover; as evidence for it, for
the same YBCO system, neutron scattering (occurrence of
a new magnetic phase) data points and two points from a
kink in ultrasound velocity are shown in Fig. \ref{fig9} from a paper by Shekhter et
al. \cite{Shekhter2013}.

The same figure also shows the onset of Kerr
rotation, which occurs at a distinctly lower temperature,
and is more compatible with the idea of a crossover. The
controversy is unresolved.

Two widely prevalent conventional
views about the pseudogap state are that it is either a state
with preformed Cooper pairs, namely with Cooper pairs
which are present but not phase coherent, or that it is a
state with competing orders, i.e. that other kinds of
electronic correlations compete with superconductive
ordering tendency. There are indeed several such
correlations such as stripes, charge density wave order,
spin order, electronic nematic….\\

\section{Strange metal:resistance and magnetoresistance \label{sec-IV}}

The metallic state beyond the pseudogap line (generally with
doping $x>x_{opt})$, quite actively investigated recently, is strange. It does not seem to have well defined electronic quasiparticles. The dc electrical resistivity and magnetoresistance are linear in temperature and magnetic field respectively. We discuss this state now.

\begin{figure}[ht!]
\includegraphics[width=\linewidth]{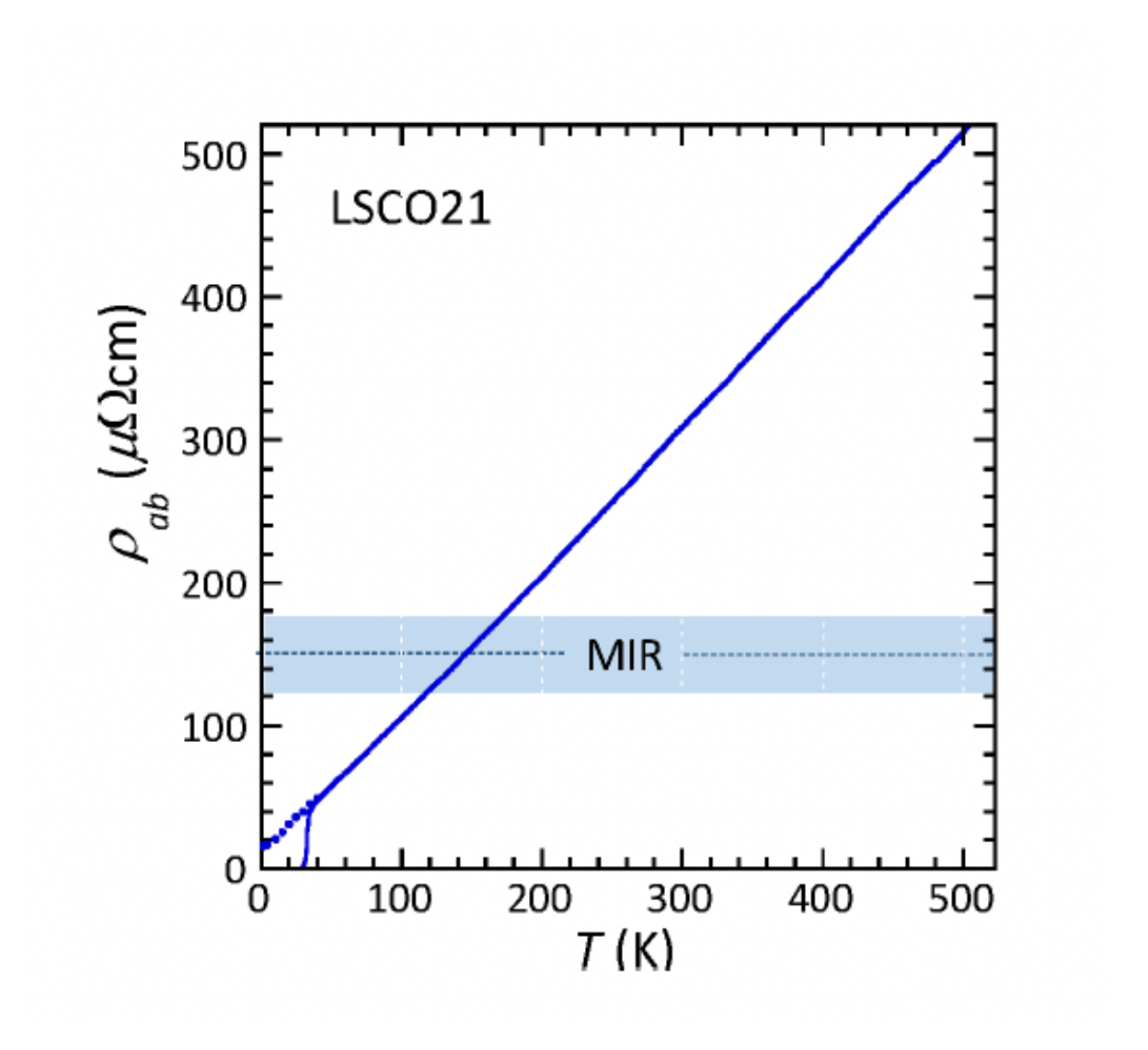}
\caption{Electrical resistivity of $La_{2-x}Sr_xCuO_4$ for $x=0.21$, $ab$ plane as a
function of $T$ (from Phillips, Hussey and Abbamonte , \cite{Phillips2022}).}
\label{fig10}
\end{figure}

A figure from a recent article by
Phillips, Hussey and Abbamonte \cite{Phillips2022} entitled ‘stranger than metals’
highlights the electrical resistivity. 
It shows the resistivity of slightly overdoped $LSCO$, with $x=0.21$.
It increases linearly with temperature from $T_c(\approx 30K)$ to the highest accessible
temperature (in this case about $500K$). \footnote{Linear
resistivity has been recognized for long as a basic pointer
towards a possibly new kind of quantum electronic state, e.g.
by Anderson $(1987)$ (Baskaran, private communication).} There
are no signs of its bending over, namely of of resistivity saturation
(which is a common but not well understood phenomenon
in many highly resistive clean metals, see e.g. ref. \cite{Gunnarsson2003}). The
resistivity goes right through the quantum limiting value, called the Mott Ioffe Regel or MIR limit (in this case,
estimated to be about $150 \mu \Omega cm$) and can be much larger than
it .\footnote{The quantum limit is estimated via
the minimum mean free path condition for resistive scattering of an
electron at the Fermi energy. The Mott limit corresponds to
assuming that it is equal to the quantum or de Broglie
wavelength of that electron, and the Ioffe-Regel limit to taking it
to be the average interatomic spacing. For typical metallic
electron densities, the values for either limit are quite close to
each other, and lead to numbers in the range of $150-400 \mu \Omega
cm$.} The resistivity below $T_c$ is obtained by destroying
superconductivity with a large enough magnetic field $B$ applied
perpendicular to the $ab$ plane, and extrapolating the observed
resistivity $\rho(T,B)$ to that for $B=0$ (this is most commonly done
by successfully fitting the measured $\rho(T,B)$ to an empirical
formula and taking $B$ to zero in it). The quantity $\rho(T,0)$ is
shown as a set of blue dots. We see that that the blue dotted
line is linear in $T$ and has the same slope. Thus, virtually from
$T=0$ to the highest temperature, the electrical resistivity is
linear.

\begin{figure}[ht!]
\includegraphics[width=\linewidth]{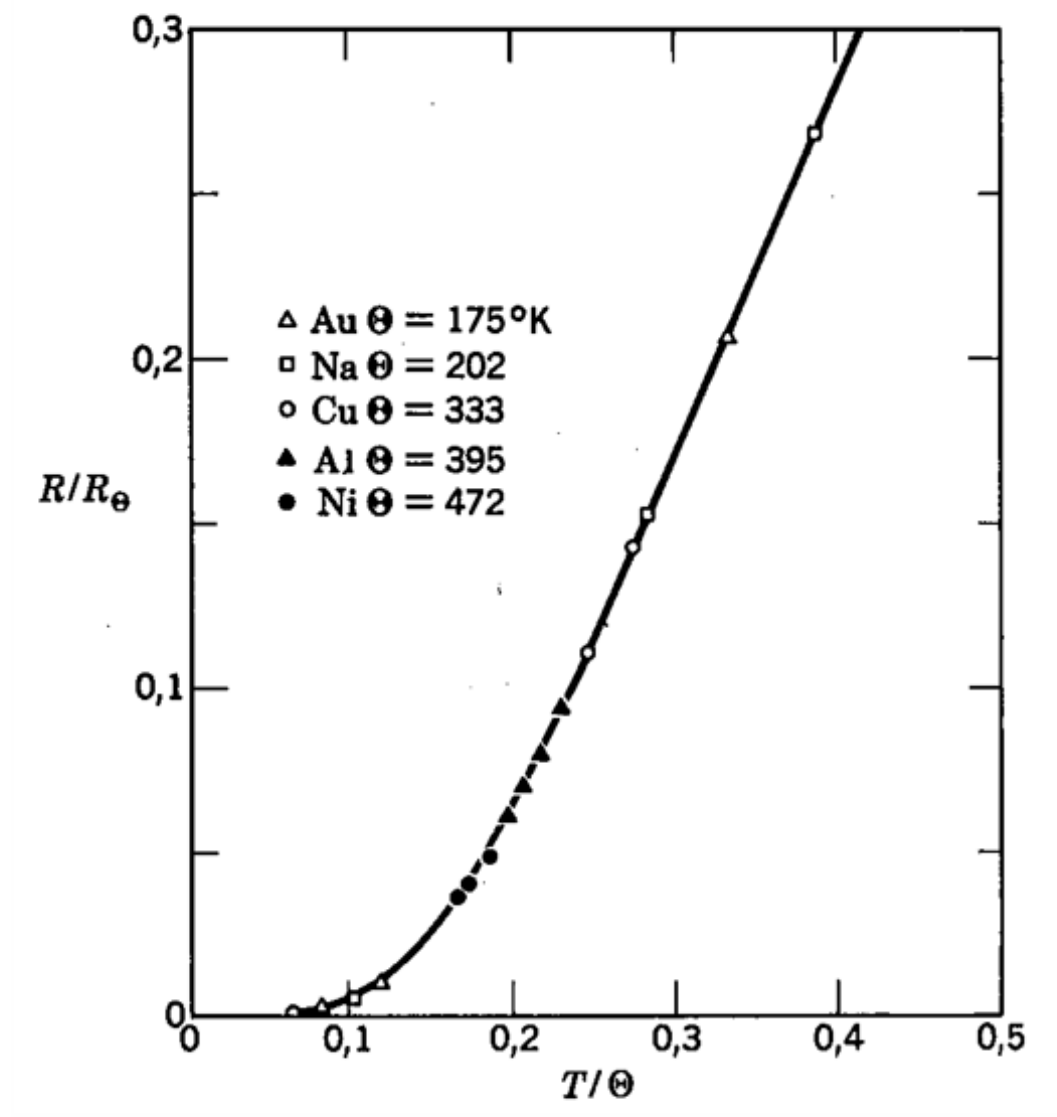}
\caption{Electrical resistivity of some clean metallic elements as a
function of $T$.}
\label{fig11}
\end{figure}

To ‘set the stage’ , we show in Fig. \ref{fig11} the resistivity of some
metallic elements as a function of temperature. The resistivity is
due to the scattering of electrons by lattice vibrations which
have a characteristic quantum scale, namely the Debye temperature $\theta_D$
(symbolized by $\theta$ here). The dimensionless resistivity ($\rho(T)/\rho(\theta))$
is plotted as a function of temperature $T$ in units of $\theta$, namely
$(T/\theta)$, and is seen to be nearly identical for all of them, though $\theta$
varies from $175K$ to $470K$ . The well known Bloch-Grueneisen
formula for electron phonon resistivity (full line) fits them. We
notice that for $T < 0.25\theta$, the resistivity is sublinear because of
quantum effects; it is linear for higher temperatures. By
contrast, the resistivity of cuprates is linear from the lowest
temperatures. 

\begin{figure}[ht!]
\includegraphics[width=\linewidth]{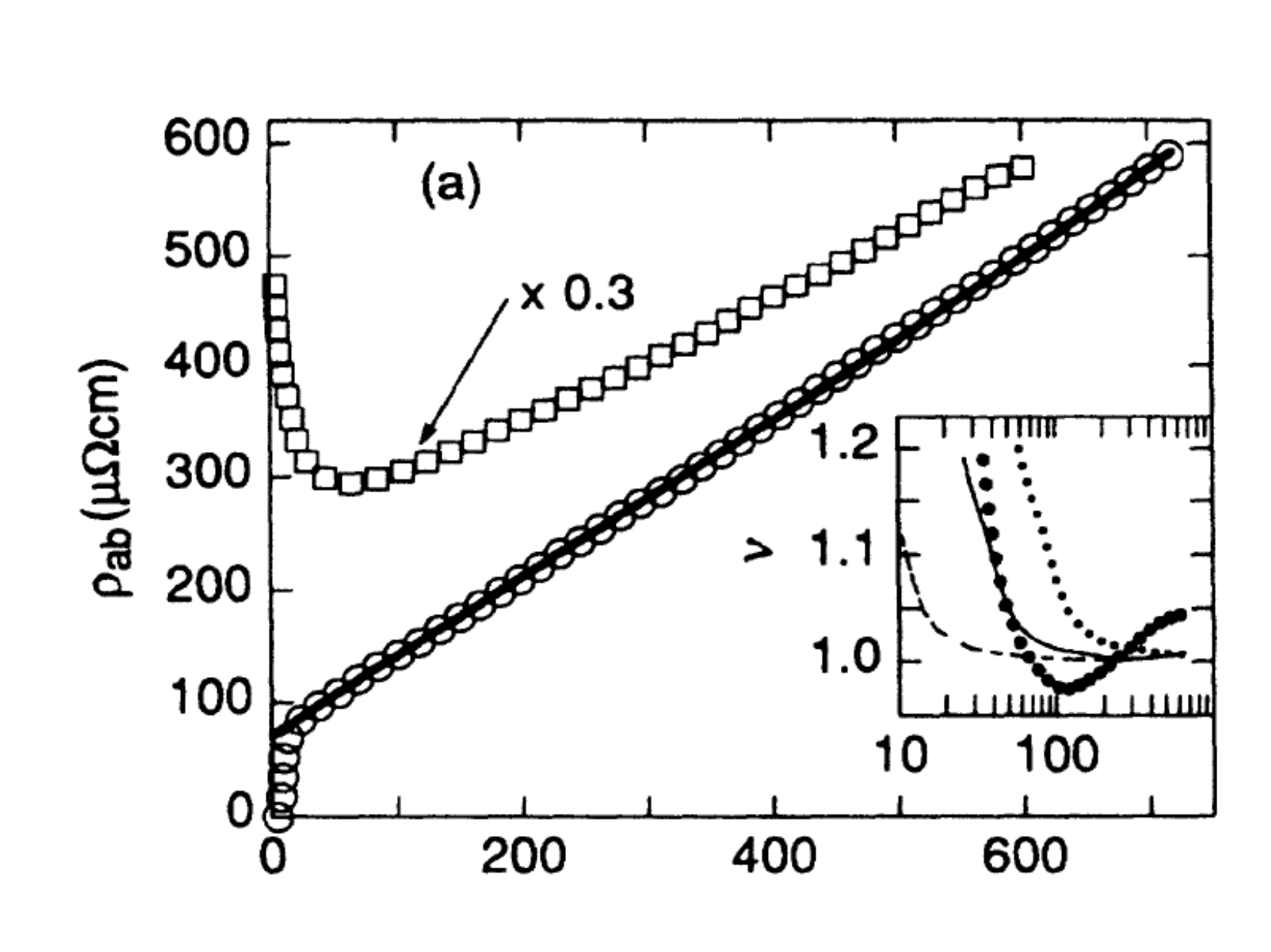}
\caption{Electrical resistivity of a single crystalline flake of $Bi-2201$, $ab$
plane (from Martin et al., ref. \cite{Martin1990}).}
\label{fig12}
\end{figure}
\begin{figure}[ht!]
\includegraphics[width=\linewidth]{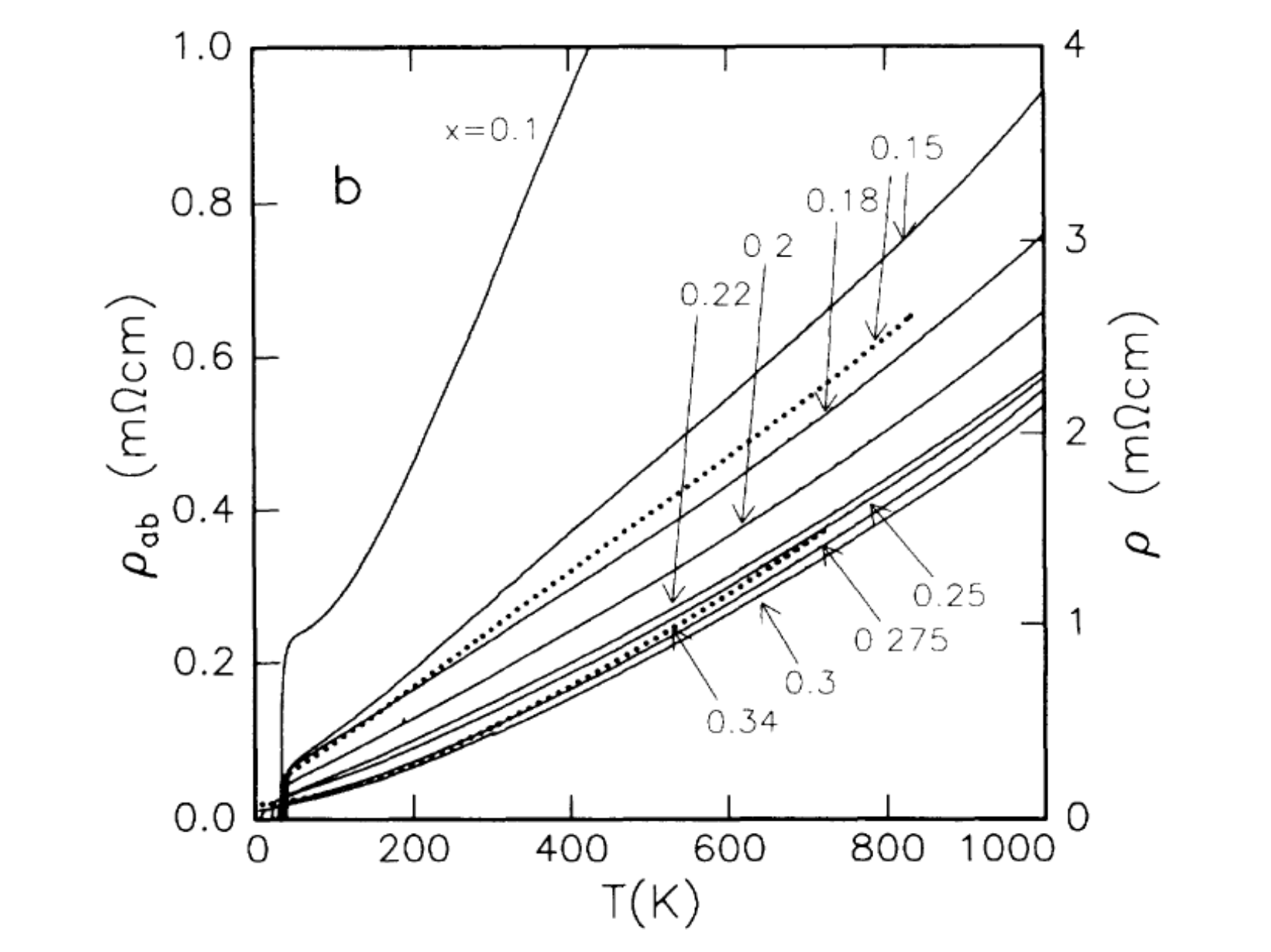}
\caption{Electrical resistivity of hole doped $LSCO$ for a wide range of hole doping and temperature (from Takagi et al., ref. \cite{Takagi1992})}
\label{fig13}
\end{figure}

An early result for the $ab$ plane resistivity of single crystalline
flakes of $Bi-2201$, due to Martin et al. \cite{Martin1990} which revealed this is shown in
Fig. \ref{fig12}.
They observed the resistivity to be
linear in $T$ from $T_c$ $(\approx 7K)$ to $700K$. Attempts to fit
the observation to the Bloch Grueneisen formula are shown in
the inset; the set of full black dots is the observed resistivity
(actually what is shown is $\nu(T)$ assuming that $\rho$
varies as $T^{\nu}$ at any particular temperature $T$). None of the
curves with various presumed Debye temperatures varying
from $10K$ to $80K$ fits the data well, even though these are unacceptably low values since the
thermodynamic Debye temperature is expected to be $350K$. Another early result, the 
resistivity of $LSCO$ for a huge range of hole doping levels from
$x=0.1$ to $x=0.34$, namely underdoped to highly overdoped, is
shown in Fig. \ref{fig13} (from ref. \cite{Takagi1992} ). 

\begin{figure}[ht!]
\includegraphics[width=\linewidth]{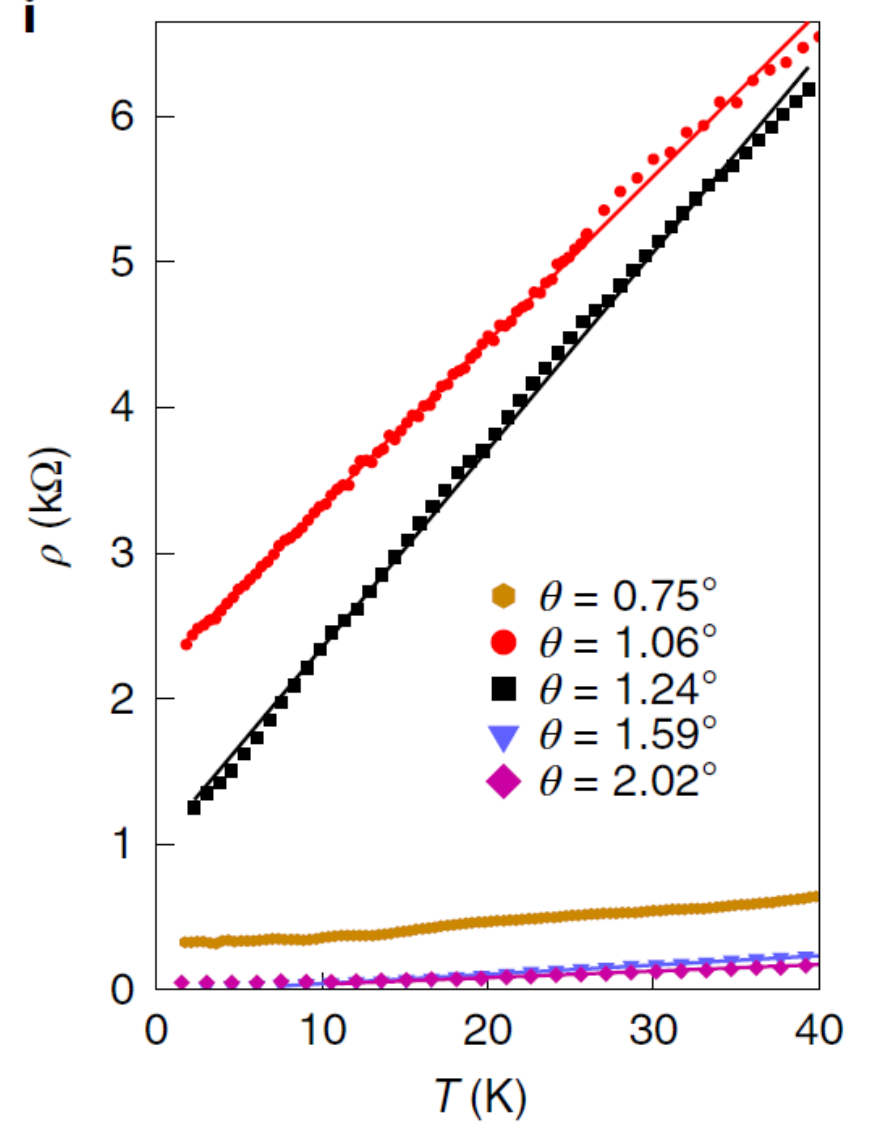}
\caption{Electrical resistivity (resistance per square) of twisted bilayer
graphene for different twist angles near the magic twist angle ( from
Polshyn et al., ref. \cite{Polshyn2019}).}
\label{fig14}
\end{figure}

The MIR limit is about $0.4m \Omega cm$. We see very nearly linear
resistivity over this entire doping range, upto high temperatures
of order $1000K$, with values well above the MIR limit. In twisted bilayer graphene, for example, the effective upper temperature can be
extended by orders of magnitude; the resistivity is seen to continue being linear. This is achieved in twisted bilayer graphene (TBG) for angles near the magic angle
of ${1.1}^\circ$, for which the tight binding graphene band is flat ($t_{eff} =0$) as follows. The bandwidth is
specially small close to this twist angle, on both sides of it.
The local electron repulsion is not affected seriously by small
changes in the twist angle between the bilayer constituents, so
that one will inevitably have $(U/t_{eff}) >>1$ or effectively strong correlation very close to the magic
angle; the ratio obviously decreases as one moves away from
the magic angle (and $t_{eff}$ increases). If linear resistivity is a
characteristic feature, not just of cuprates, but of all strongly
correlated systems, one expects linear resistivity near the
magic angle in TBG as well. Further, because of the small
carrier density in these systems (typically, $n_h \approx 10 ^{12}/cm^2)$ the
effective Fermi energy $\varepsilon_F^*$  is small $\approx 30-35K$ and one can 
investigate the resistivity upto temperatures of the order of $\varepsilon_F^*$.
The resistivity of $TBG$ at different angles of twist and
temperatures is shown in Fig. \ref{fig14} ref. \cite{Polshyn2019}. We see that a nearly linear temperature
dependence of resistance continues upto temperatures as high
as $\varepsilon_F^*$. (In the cuprate language, we can access temperatures as high as $\sim 20K$ !).

Things have got curioser in the last few years, with high quality measurements and analyses of $\rho(B,T)$ for clean, well characterized, mostly overdoped, cuprates ; the external field $B$
ranges here upto $\approx 60T$. The overdoped regime is preferred
because the complexities of pseudogap and other possible
competing order such as charge order (CDW?) are absent
there. Further, since $T_c$ is small, it is possible to destroy
superconductivity with large but accessible magnetic fields. 

\begin{figure}[ht!]
\includegraphics[width=\linewidth]{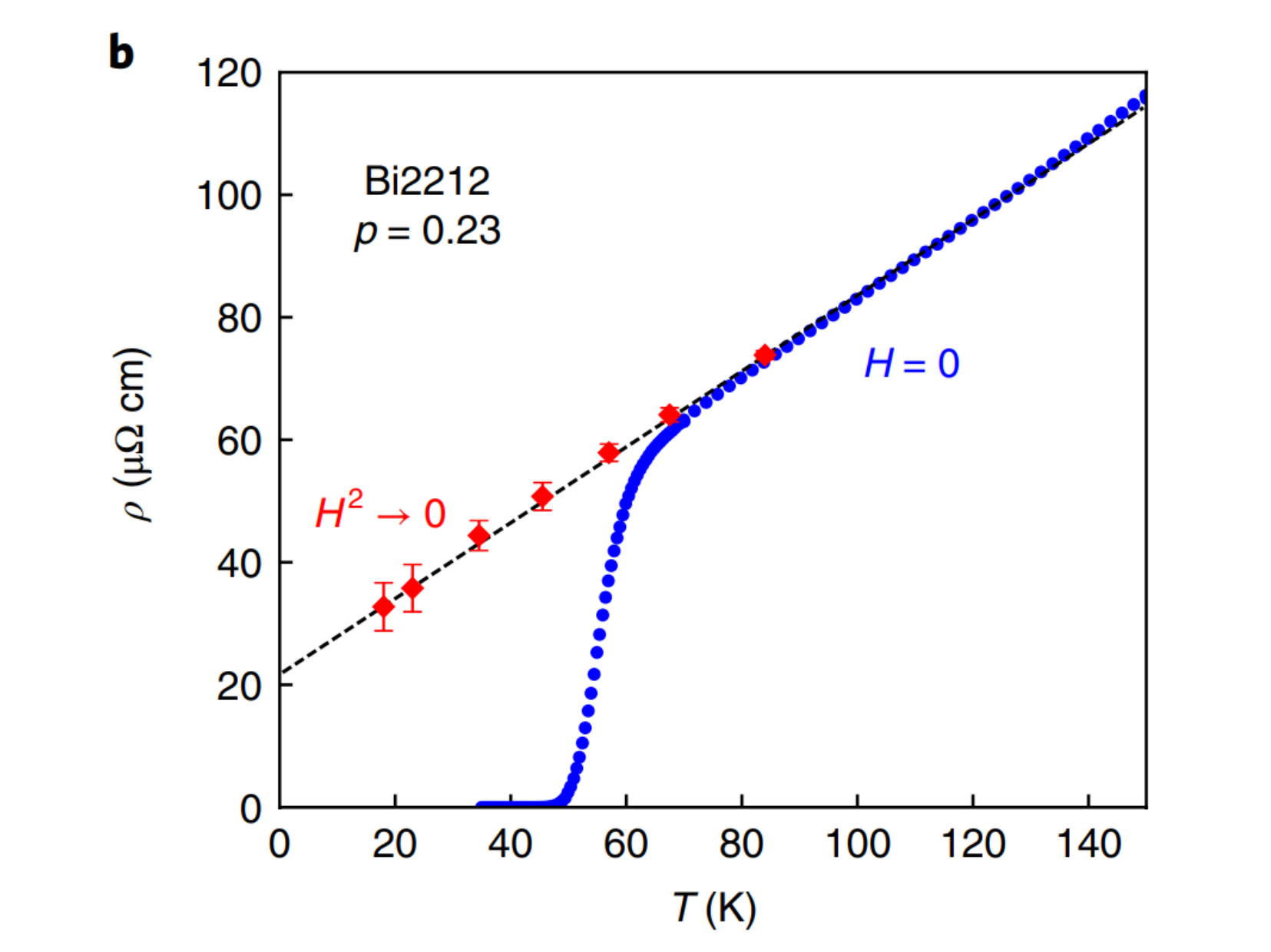}
\caption{Electrical resistivity of overdoped $Bi-2212(p=0.23)$,
extrapolated to $B^2 =0$ (from Legros et al., ref. \cite{Legros2019}).}
\label{fig15}
\end{figure}

\begin{figure}[ht!]
\includegraphics[width=\linewidth]{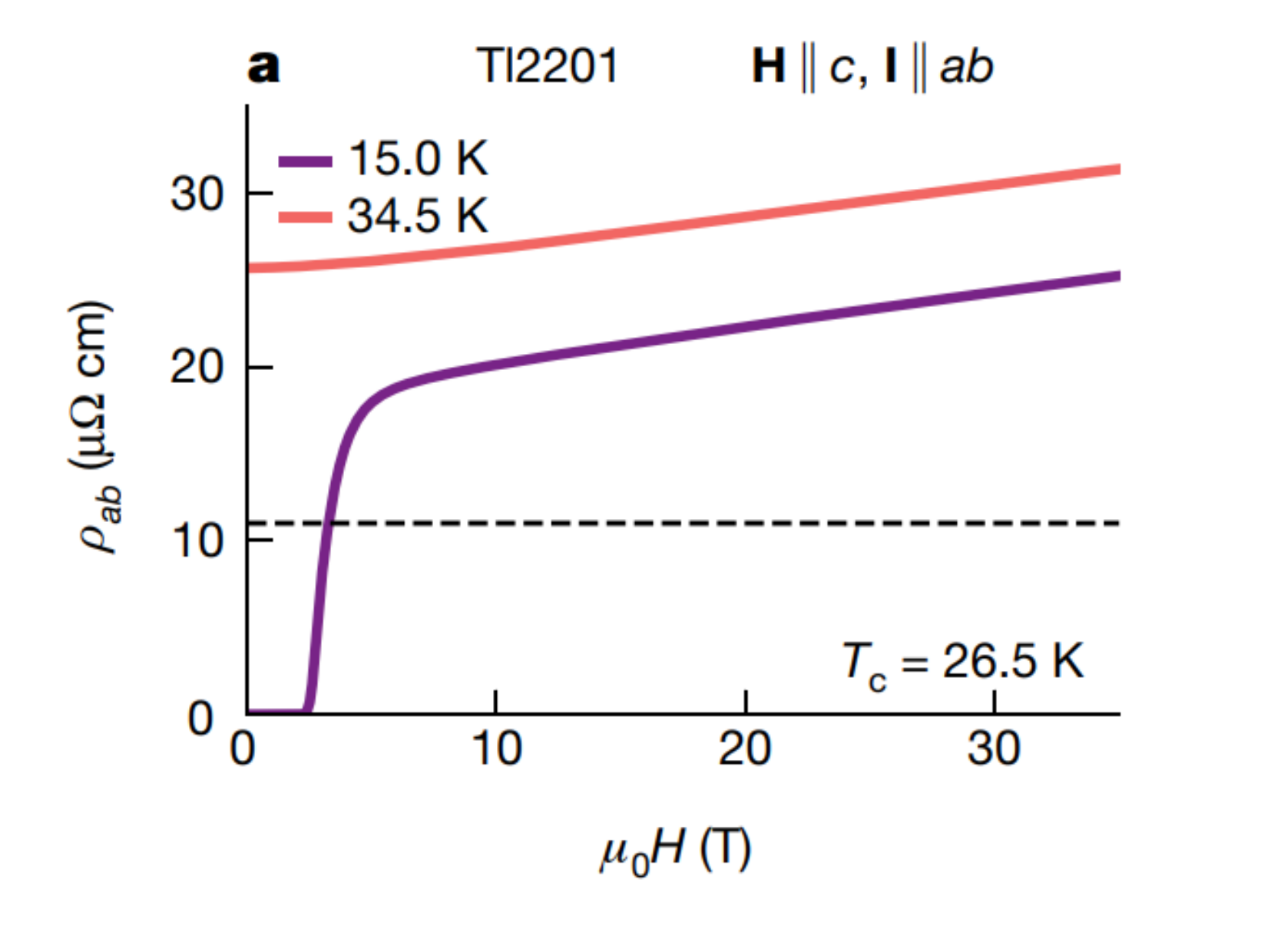}
\caption{Linear magnetoresistance in overdoped $Tl2201(p=0.27)$ (from
Ayres et al., ref. \cite{Ayres2021}).}
\label{fig16}
\end{figure}

An example of the results is shown in Fig. \ref{fig15} taken from the paper of Legros et al. \cite{Legros2019}. The
magnetic field dependent part of $\rho(B,T)$ is seen to go as $B^2$ for
different small values of $B$. This enables one to extrapolate to
$\rho (0,T)$. Because the measurements are made in high fields,
one can access the regime $T <T_c$ , and find $\rho(0,T)$ for $T<T_c$.
Some results for overdoped $Bi-2212$ $(x=0.23)$ are shown in the
figure. We see that the data fall on the linear resistivity curve for
$T >T_c$ extrapolated to values below $T_c$; the resistivity slopes are the same. This is quite remarkable,
also because it is a physical property of a thermodynamically
unstable state (the stable state below $T_c$, at $B=0$, is of course
superconducting with $\rho(T) = 0!$). Even more interestingly, the
magnetoresistance at high fields is observed to be linear in $B$.
This is seen from Fig. \ref{fig16}.
(Ayres et al., ref. \cite{Ayres2021}) which shows measurements on
overdoped $Tl-2201$ (with $x\approx 0.27$ and $T_c \approx 26.5K$). 

\begin{figure}[ht!]
\includegraphics[width=\linewidth]{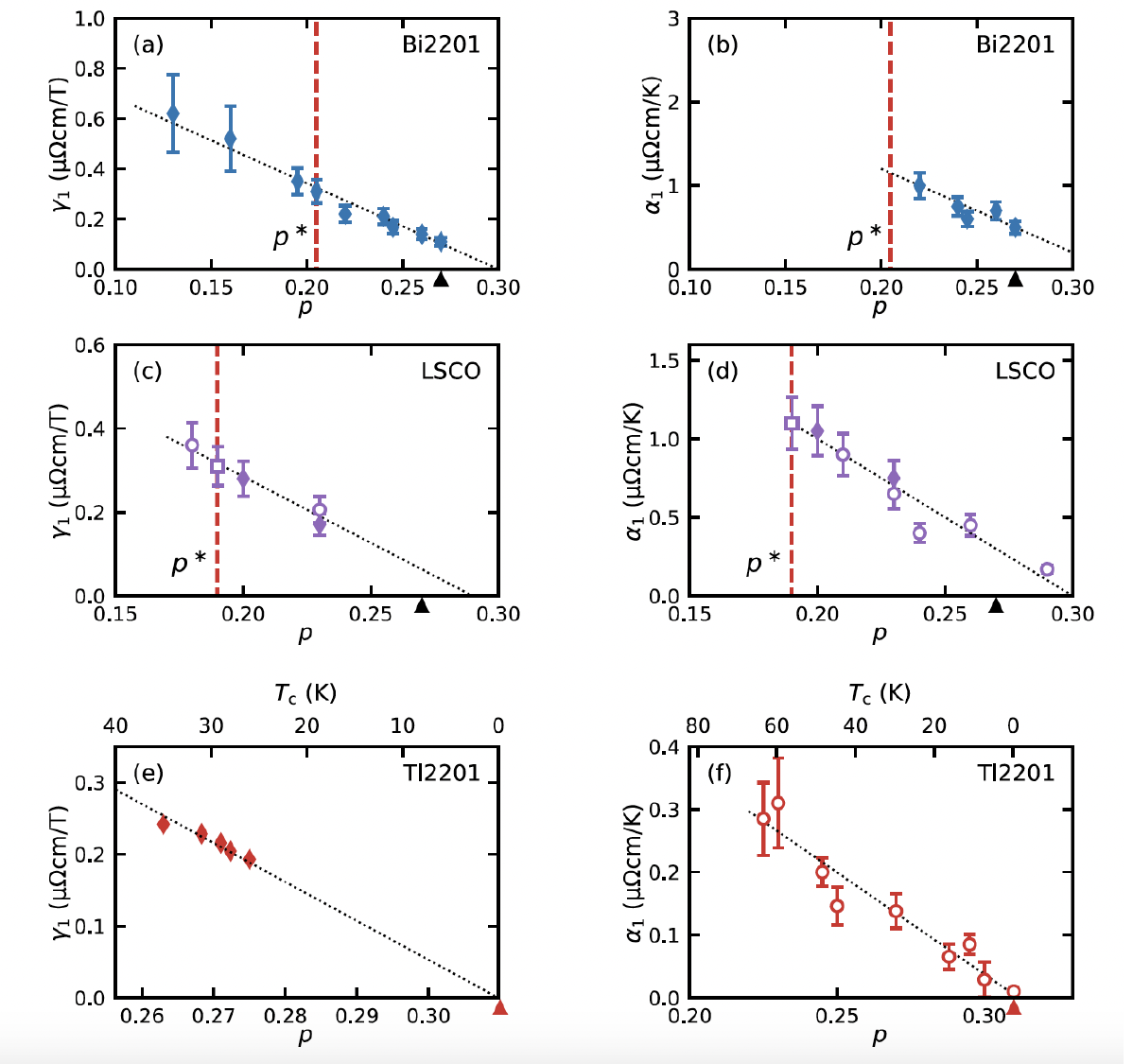}
\caption{Comparison of slopes of linear resistivity and linear
magnetoresistance in three overdoped cuprates (from Ayres et al.,
ref. \cite{Ayres2024}).}
\label{fig17}
\end{figure}

This is surprising since so long as $\mu_BB <<\varepsilon_F$ or $\mu_BB <<\hbar/\tau$, which is
generally the case in metals, the magnetoresistance goes as
$B^2$. Here however, the magnetoresistance starts as $B^2$ for low
fields and becomes proportional to $B$ for high fields such that $\mu_BB>k_BT$.
The slope of the magnetoresistance is seen to be independent
of $T$. The empirical relation $\rho(B,T)= F(T) +\sqrt{(aT)^2 + (bB)^2}$ is
found to fit the data well. (This form has of course the observed low and high $B$ behaviour). Even more intriguing is the recent finding that in energy units, the slope of the linear resistivity and of the
linear magnetoresistance are the same for the three overdoped
cuprates investigated (ref. \cite{Ayres2024}, Fig. \ref{fig17}). 

\section{Empirical correlations \label{sec-V}}

\begin{figure}[ht!]
\includegraphics[width=\linewidth]{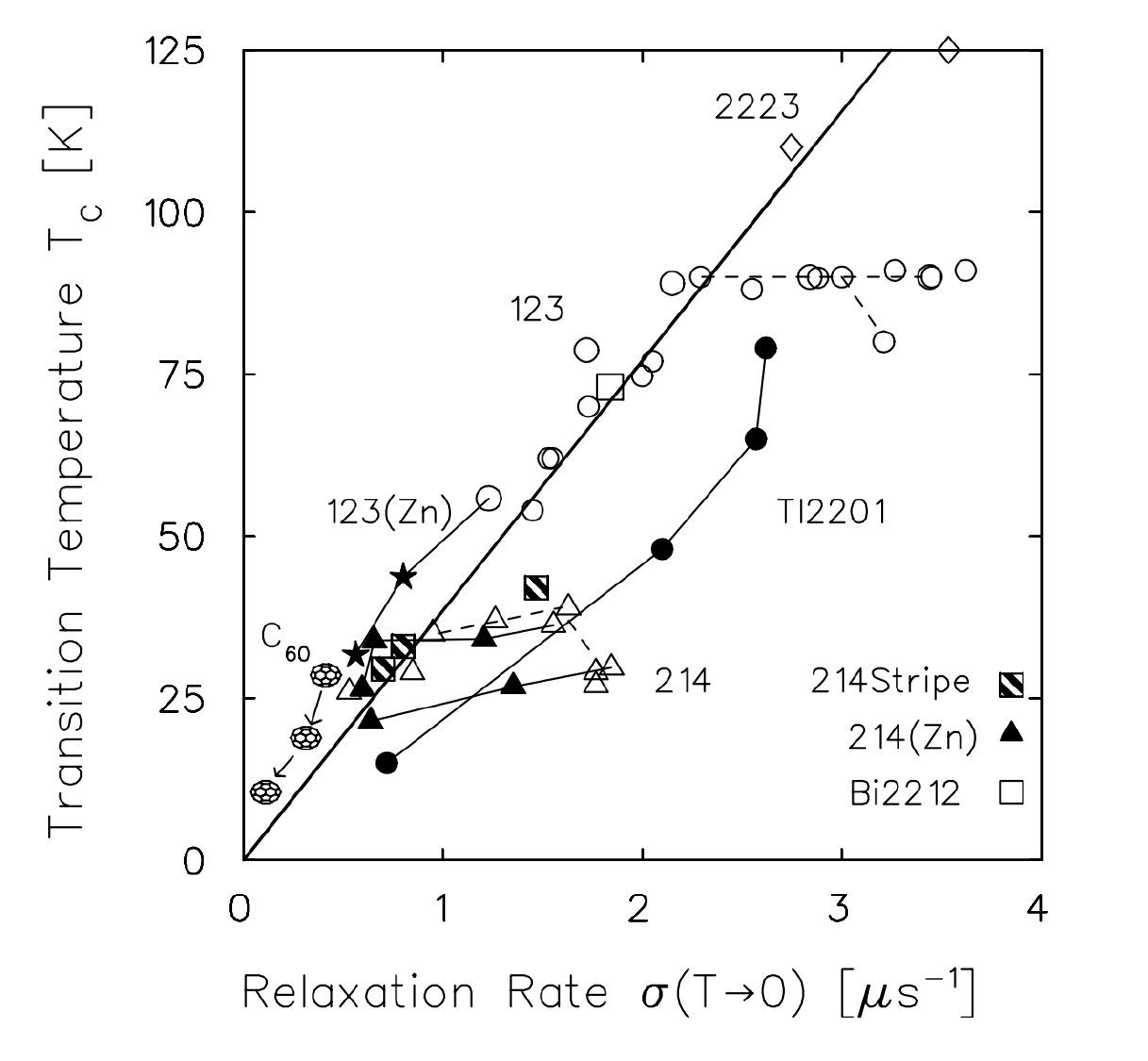}
\caption{Uemura correlation between $T_c$ and superfluid density (measured here via muon spin relaxation) (from Uemura, ref. \cite{Uemura2004}.}
\label{fig18}
\end{figure}

\begin{figure}[ht!]
\includegraphics[width=\linewidth]{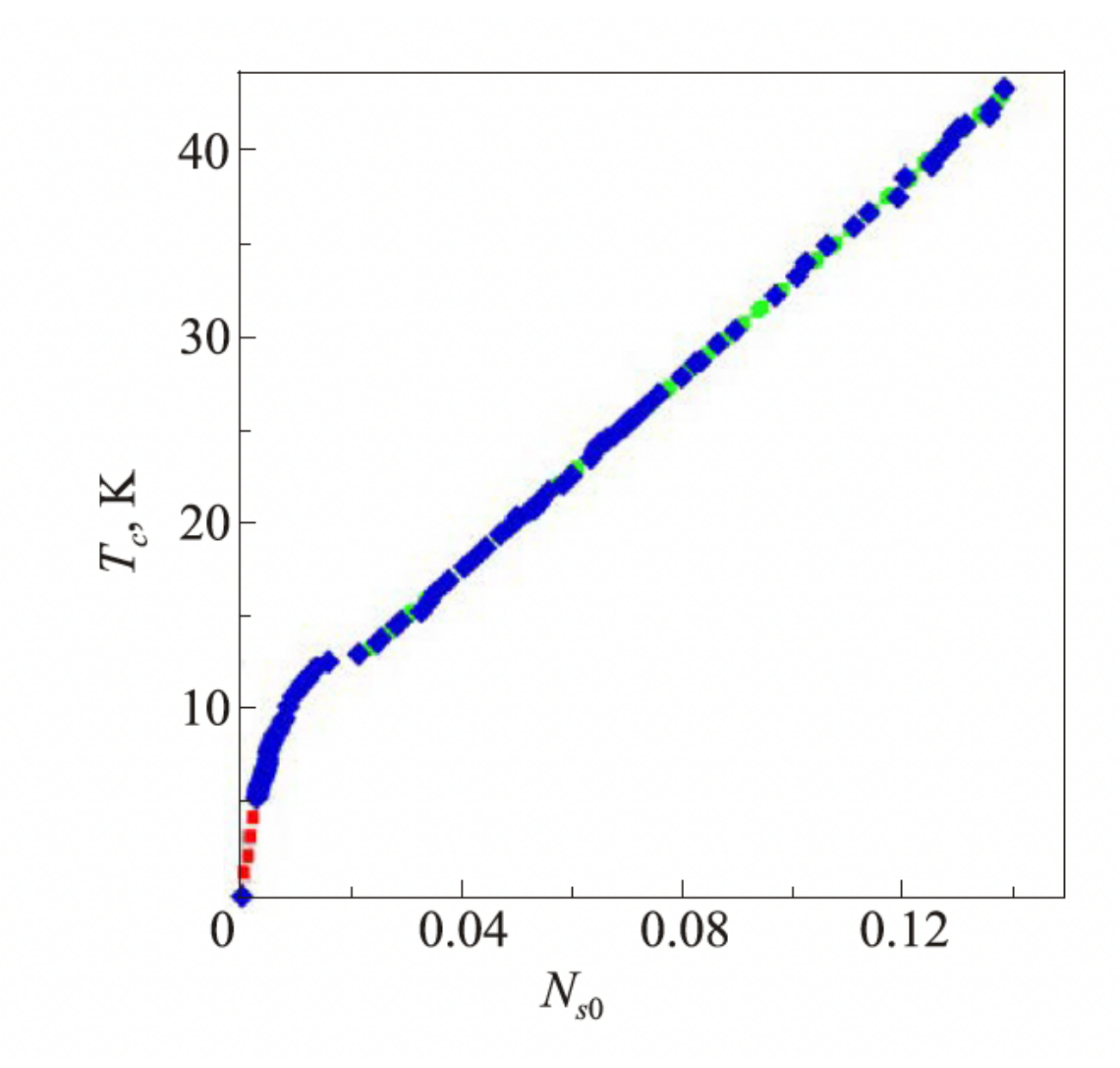}
\caption{Correlation between $T_c$ and superfluid stiffness in overdoped
$LSCO$ (from Bozovic et al., ref. \cite{Bovzovic2016}).}
\label{fig19}
\end{figure}

We present below some empirically observed correlations
between two experimentally measured cuprate
superconductor quantities.
Perhaps the oldest such correlation, due to Uemura
$(1989)$, is between two equilibrium properties of the superconductor, namely between $T_c$ and superfluid density or more
accurately $T_c$ and $(1/\lambda^2)$ where $\lambda$ is the London or magnetic
penetration depth.\footnote{Uemura measured the width $\sigma$ of
the Gaussian spread in the Larmor precession frequencies of
decaying spin polarized muons injected into the cuprate
superconductor. A magnetic field is applied perpendicular to
the ab plane. He found that the width  is proportional to $T_c$, with a universal slope. One can show fairly generally that $\sigma \propto 1/\lambda^2$ , so that one has $T_c \propto 1/\lambda^2$.} A figure from a more recent paper \cite{Uemura2004} by Uemura
(Fig. \ref{fig18} shows a
broad correlation for a number of cuprates (generally in the
underdoped regime) and other systems.
The correlation is
not restricted to underdoped cuprates; for example, it is
observed in well characterized overdoped $LSCO$ films with $x$
ranging from about $0.17$ to $0.27$. Here, the absolute value of $1/\lambda^2$ was measured accurately by a mutual inductance
method. The connected superfluid stiffness $N_{so}(=(A/\lambda^2)$ where $A$
involves only universal constants) displayed against $T_c$ \cite{Bovzovic2016}
(Fig. \ref{fig19}).

A clear linearity is
evident. Very near the $QCP$ where $T_c \rightarrow 0$, the quantum
critical behaviour $\sqrt N_{so} \propto T_c$ is observed. A peculiar common feature of the superconducting state $(T <<T_c )$, perhaps related to this, is that $N_s \propto T$ over
a wide range of temperatures except for very low ones,
where it goes as $T^2$.

\begin{figure}[ht!]
\includegraphics[width=\linewidth]{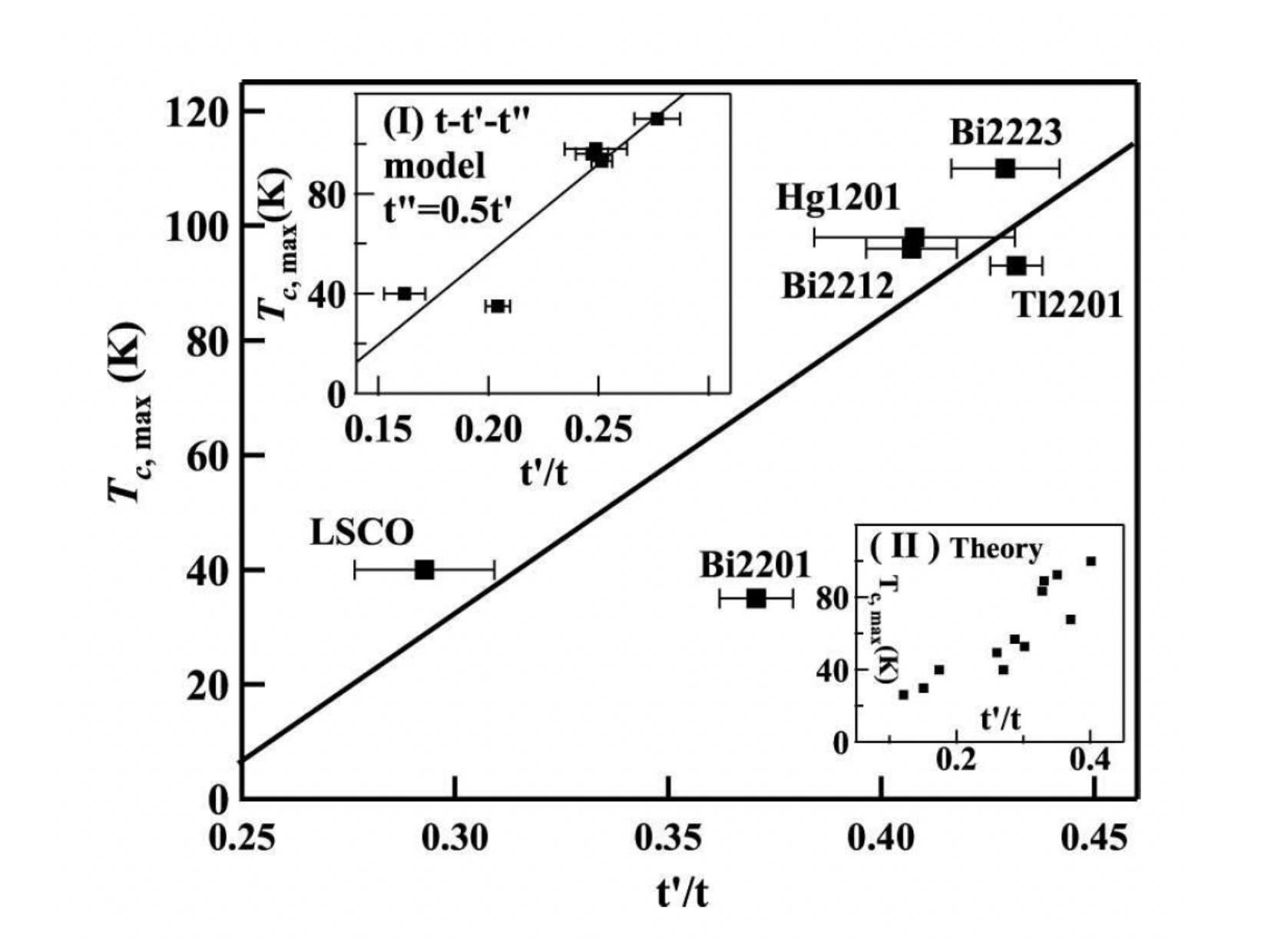}
\caption{Correlation between the maximum $T_c$ of a cuprate, namely
$T_{c,max}$ and next nearest neighbour hopping $t^\prime$ (in units of nearest
neighbour hopping $t$) (from Lee et al. ref. \cite{Lee2006}).}
\label{fig.20}
\end{figure}

An older empirical correlation is between $(t^\prime/t)$ and $T_{c,max}$ due to Pavarini et al. \cite{Pavarini2001}. An effective one band model for a
stoichiometric cuprate family member was obtained by
starting with a large number of possibly relevant states and integrating out states 
other than those corresponding to the tight binding local $d_{x ^2 - y^2}$
symmetry band. The ’renormalized’ parameters of the
effective one band model were determined. It was noticed,
surprisingly, that $(t^\prime/t)$ ( obtained for the undoped cuprate) is proportional to the maximum value of
$T_c$ of the hole doped cuprate. A purely experimental version of this
correlation, for the actual hole doped cuprate is due to W. S. Lee et al. \cite{Lee2006}. They determined the
location of the Fermi surface of the metal with $T_{c,max}$ from
ARPES and fitted it with tight binding parameters. Their plot
of $T_{c,max}$ vs. $(t^\prime/t)$ (inferred at $x_{opt}$ ) is shown in Fig. \ref{fig.20}.

\begin{figure}[ht!]
\includegraphics[width=\linewidth]{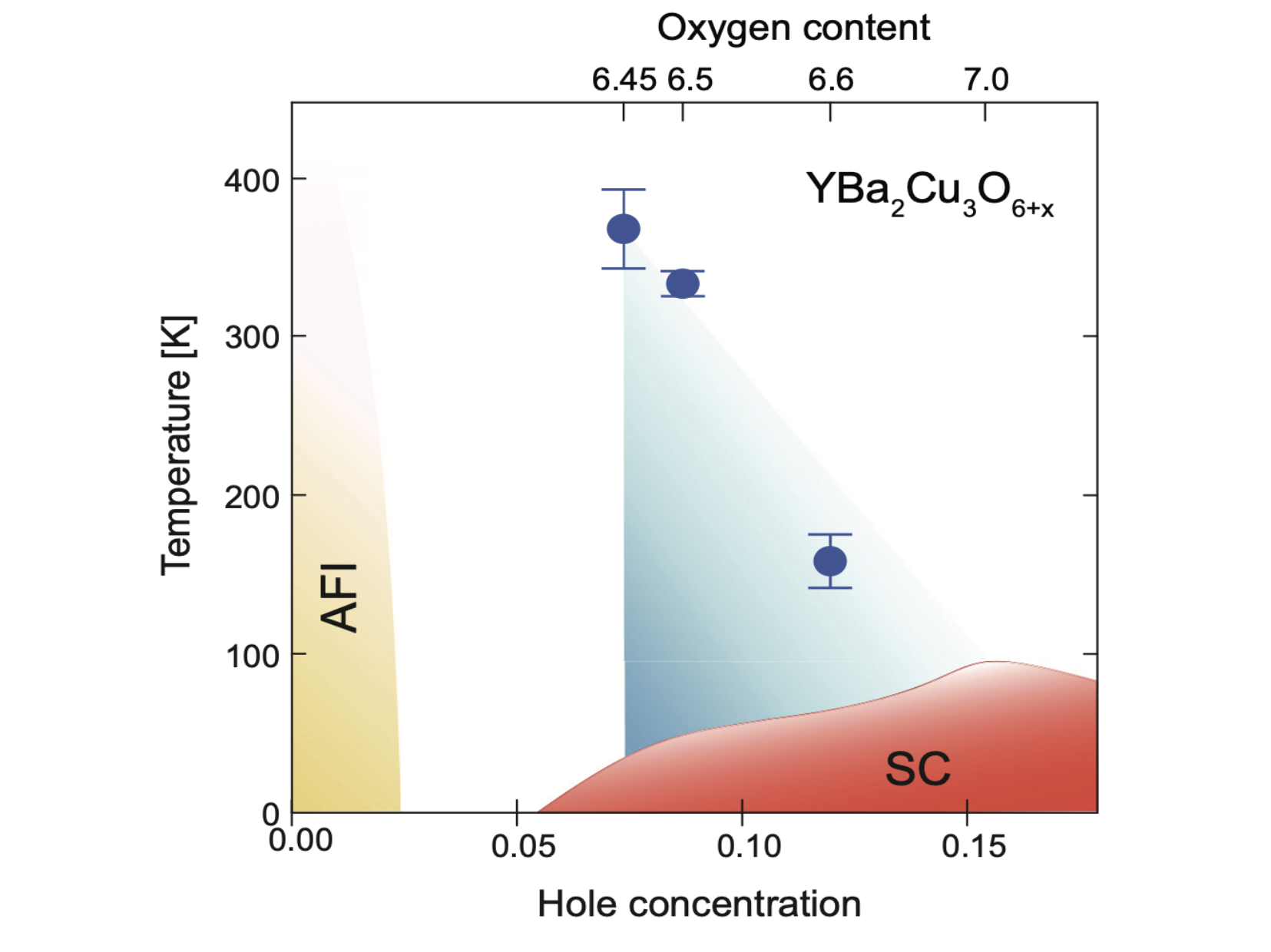}
\caption{Correlation between nonequilibrium superconducting phase
generated by resonant laser excitation of apical oxygen vibrational
mode and hole density in $YBCO$, in the temperature $(T)$ hole density
($x$ or $p$) plane (from Kaiser et al. ref. \cite{Kaiser2014}).}
\label{fig21}
\end{figure}

This is surprising, because
the superexchange term $J\textbf{S}_i.\textbf{S}_j$ where
$J\approx(t^2/U)$ can also be thought of as the nearest neighbour
(intersite) Cooper pairing term in the Hamiltonian; one therefore expects a linear
relation between $T_c$ and $t^2$, not between $T_c$ and $t^\prime$. Support for
$t^\prime$ being relevant comes also from huge enhancement of $d$ wave superconducting correlations (observed 
 via ac conductivity measurements) due to resonant laser excitation
of the apical oxygen vibrational mode. (Ref. \cite{Kaiser2014}, Fig. \ref{fig21})
 
One can interpret this as follows. The next nearest neighbour hopping amplitude $t^\prime$
has a large contribution from the overlap between the $p$
orbital of the apical oxygen with $d_{x^2 - y^2}$ symmetry orbital at
sites $i$. One therefore has a term in $t^\prime$ which is of the second
order in the overlap $t_{pd}$. The $pd$ overlap and so $t_{pd}$ can be
increased by resonant excitation, increasing $t^\prime$ and thus $d$
wave superconducting correlations in the spirit of this
empirical finding.

\begin{figure}[ht!]
\includegraphics[width=\linewidth]{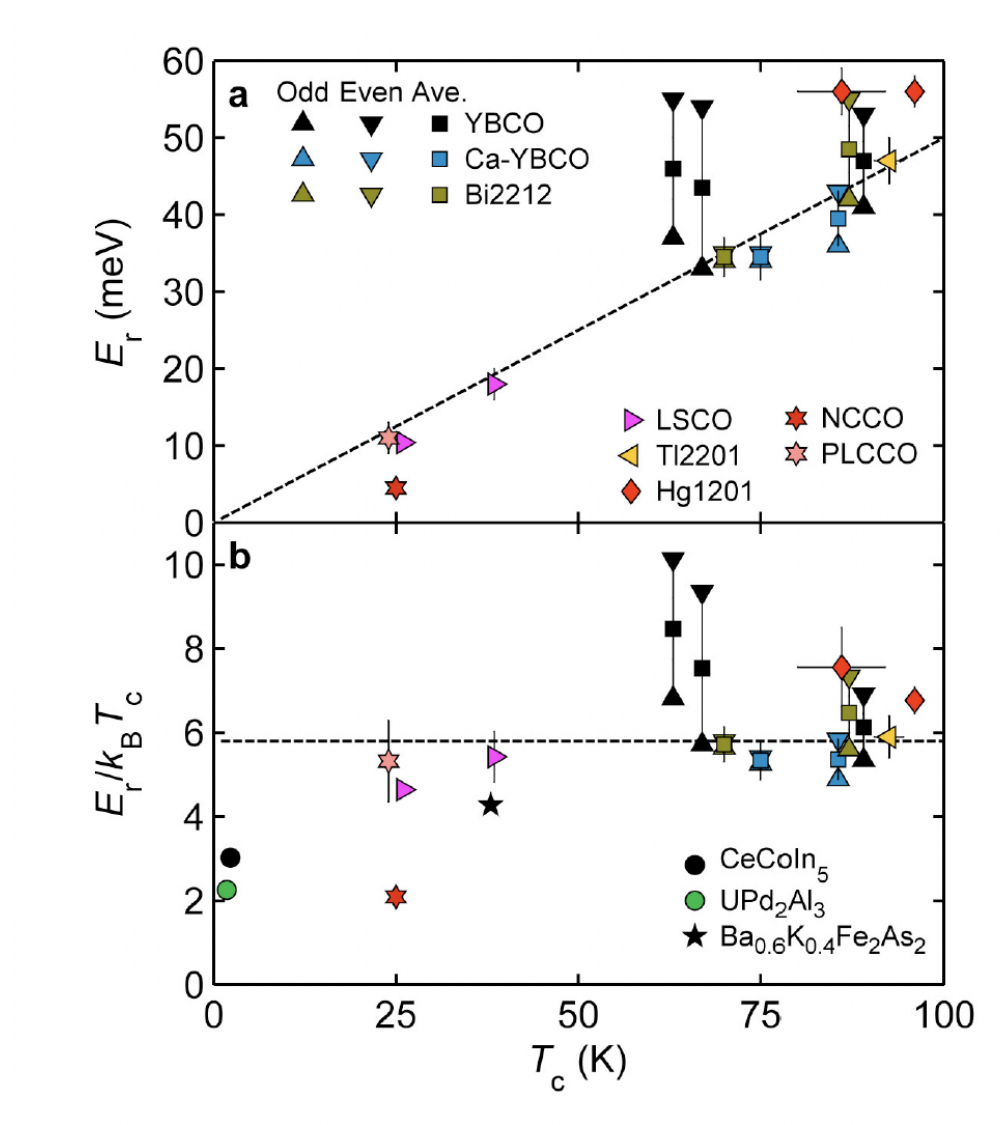}
\caption{Correlation between neutron resonance
(’$41 meV$ resonance’) energy and $T_c$ ( from Yu et al. ref. \cite{Yu2009}).}
\label{fig22}
\end{figure}

Another interesting correlation is
between a sharp peak observed in the inelastic scattering of
neutrons from the cuprate superconductor for $\textbf{Q} = (\pi/a,\pi/b)$  and $T_c$. The peak
 is quite sharp below $T_c$. (This is known as
the $’41 meV’$ resonance for historical reasons). The
resonance energy $E_r$ is proportional to $T_c$ in a number of
cuprates, with a proportionality constant close to $6$. The $T_c$
values range from about $25K$ to about $100K$ (see  Fig. \ref{fig21}). 
Finally, we mention a characteristic effect in cuprates connected with a nondissipative transport property, namely the Hall effect. It
has attracted a lot of attention recently and is the crossover
in the inverse Hall number from $p$ (hole doping) to $(1+p)$ as $p$
goes highly overdoped. This is shown in Fig. \ref{fig22} which is reproduced from a
paper by C.Putzke et al. \cite{Putzke2021}. 

\begin{figure}[ht!]
\includegraphics[width=\linewidth]{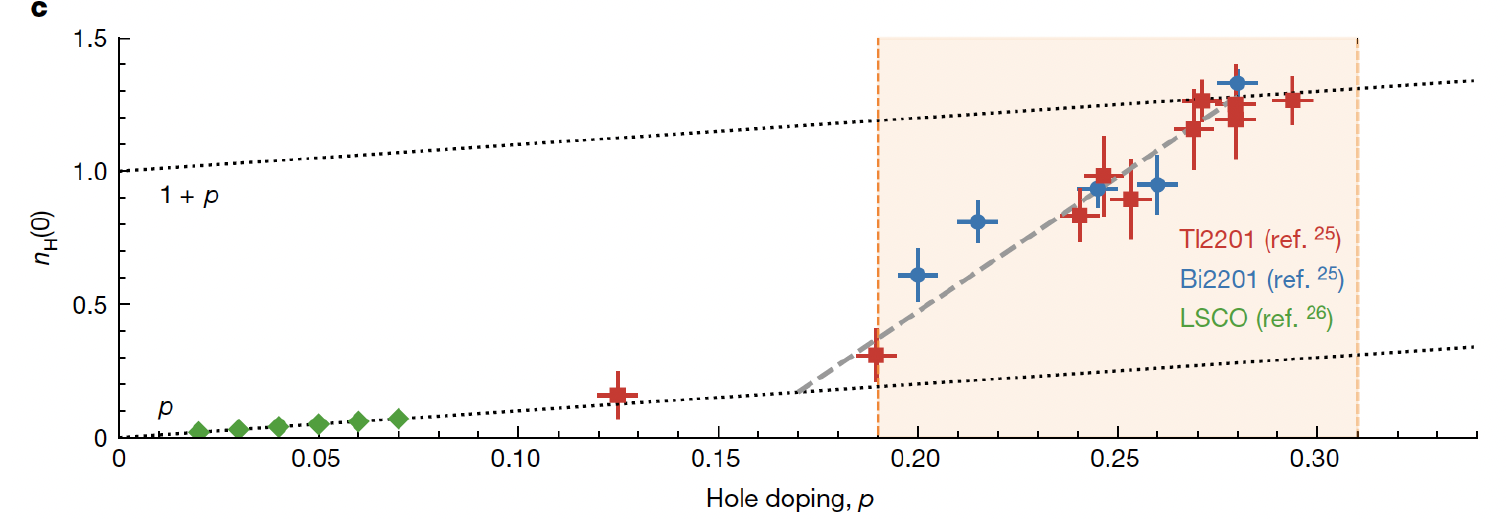}
\caption{Inverse Hall number as a function of hole doping, specially the crossover for overdoping (from Putzke et al. ref. \cite{Putzke2021}).}
\label{fig23}
\end{figure}

We see a fairly large crossover
region (from $p \approx 0.17$ to $p \approx 0.27$) over which this happens, in a
number of overdoped cuprates. As in the standard Hall
configuration, the Hall strip is in the $xy$ plane ($ab$ plane of the
cuprate). An electric field $E_x$ is applied in the $x$ direction as a
consequence of which there is an electric current in that
direction with current density $j_x$. A magnetic field $B_z$ is applied
perpendicular to the $xy$ plane (along the $c$-axis, in our case).
A Hall electric field $E_y$ develops in the $y$ direction. The Hall
number $R_H = (E_y /j_x B_z)$ and can be used to define
(experimentally) a ‘carrier density’ $n_H$ as $R_H = (1/n_H e)$. In
most metals and semiconductors, this is indeed seen to be
the density of electron or hole current carriers. In cuprates,
this number depends significantly on a temperature scale of
order $100K$, much smaller than an electronic scale, so that its
identification with carrier density is not convincing. At low
temperatures and at high magnetic fields, the experimentally
observed ratio $R_H$ rises from zero in the superconducting
state and becomes field independent as well as nonzero
when superconductivity is destroyed. The inverse of this
number (in units of $e$), namely $n_H$ is seen for $x <0.16$ to be
the hole density $x$ (it is called $p$ in the figure). \footnote {The
number of mobile electrons per unit cell is $(1-x)$ for $x$ holes
per unit cell. The total number of electron states in the
Brillouin zone is two per unit cell; for $(1-x)$ electrons in this
BZ (occupying $(1-x)$ of those states), the number of
unoccupied or hole states is $(1+x)$).} The finding is that as $x$ increases from mildly overdoped $(x > 0.17-0.18)$ to highly
overdoped ( $x \approx 0.27$), $n_H$ increases from $x$ to $(1+x)$.\\

We conclude by mentioning an unexpected relation between the
strange metal above $T_c$ and superconductivity.
The measured linear resistivity of the strange metal has a
slope $A$ which is correlated with $T_c$ in a variety of systems.
This has been known for some time (e.g. the review by
L. Taillefer \cite{Taillefer2010} in $2010$ entitled ‘Scattering and Pairing in
Cuprate Superconductors’). 

\begin{figure}[ht!]
\includegraphics[width=\linewidth]{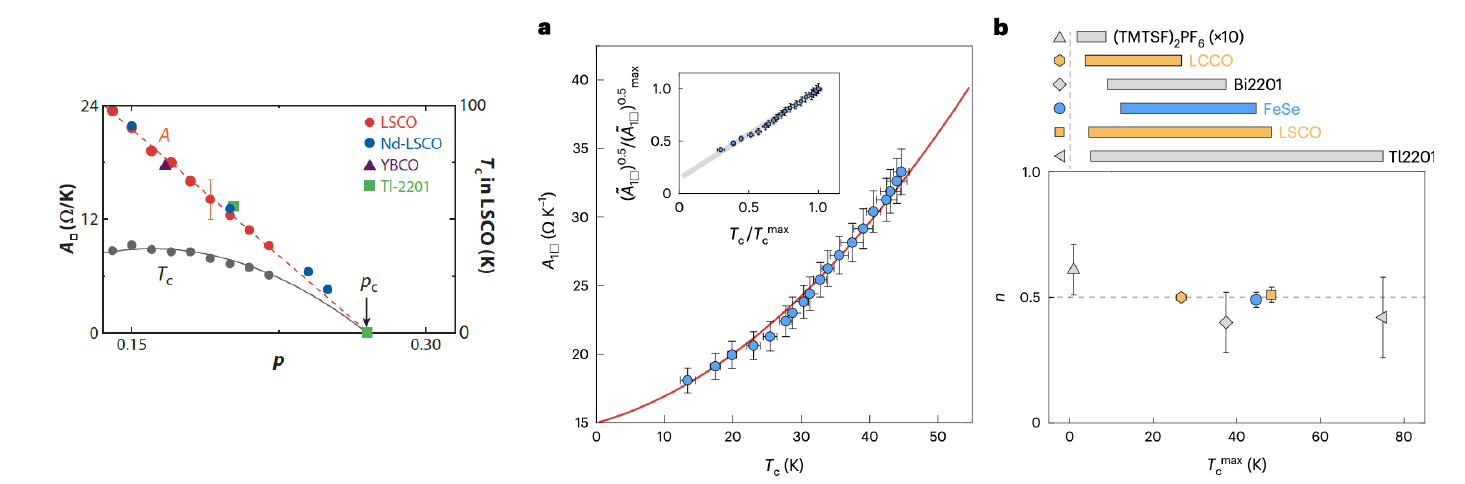}
\caption{Correlation between slope $A$ of the linear resistivity and $T_c(A^{0.5} \propto
T_c$) (from Jiang et al. ref. \cite{Jiang2023}).}
\label{fig24}
\end{figure}

More precisely, we see in
Fig. \ref{fig24} that in many compounds
(not only cuprates), $A^{0.5} \propto T_c$. The correlation was first
brought out \cite{Jiang2023} in extensive work on $FeSe$, whose $T_c$ can be
tuned from $13K$ to $45K$, and was generalized to show that it
holds for cuprates and organics in addition as is shown in
Fig. \ref{fig3} of that reference. An interesting aspect of this
correlation is that a thermodynamic equilibrium property
connected with pairing coherence, namely $T_c$ , is related to the
nonequilibrium property of electrical current, namely to electrical resistivity (which arises from incoherent or dissipaptive scattering). 
$\rho=AT$.

\section{Trying to make sense of the goings on \label{sec-VI}}

We notice that broadly, two kinds of strange properties of cuprates have been described above.  
 One is related to their superconductivity i.e. to nonzero $T_{c}$, for example the correlation between $(t'/t)$ and $T_{c}$. The other kind of behavior, e.g. linear resistivity, does not involve $T_{c}$ or superconductivity  per se. The two seem connected, at least in the cuprates;  for example, the slope $A$ of the linear resistivity is connected with $T_{c}$; specifically we have $\sqrt{A} \propto T_{c}$. The large amount of related activity generated is not discussed here; there is an overarching idea that these materials are strongly correlated, and that this is the key to all their properties including their superconductivity. If the system is very strongly correlated so that doubly occupied on site states are projected out (as is identically true for $U=\infty$), the nonsuperconducting state is a projected Fermi liquid. The idea goes back to Gutzwiller in 1963 (see for example ref.44 for a relatively recent summary) and was explored also in conjunction with the RVB mechanism of superconductivity. Its implementation over a range of correlation strengths (and ranges), doping ranges, temperatures, dimensions, lattice types etc. is an ongoing activity. As mentioned earlier, the basic ingredient of superconductivity in the strong correlation RVB approach of Anderson, Baskaran is the nearest neighbour superexchange $J$, a 'small' quantity $\sim(t^{2}/U) = t(t/U) <<t$.  

In this spirit, one should perhaps look for a new picture for many electron systems founded explicitly on the $U=\infty$ limit and build from there (for example a theory in the small parameter $(t/U)\sim (0.1)<<1$ in the cuprates) in the hope of a comprehensive understanding of this collection of strange properties. For many electron systems in a lattice model, this can be motivated by the fact that there are two obvious limits, namely one of $U=0$ and the other of $U=\infty$. The first is exactly soluble analytically for the continuum case and is described by the Drude free electron gas theory. \footnote {As is well known, independent electrons in a periodic lattice or Bloch electrons are not an exactly soluble system, but are computationally accessible. They also have novel topological properties.} For nonzero interaction (nozero $U$), there are very sophisticated and highly developed theories which are finally perturbative in $U$; they are adiabatically connected to the no interaction limit. Our efforts so far to understand the cuprates are generally rooted in this limit (often with additional auxiliary fields and constraints).  The other limit is the metallic state for infinitely strongly correlations $(U=\infty)$. The many electron problem is not exactly soluble in this limit and there are very few theoretical methods available for exploring it. We need reliable (necessarily approximate) solutions to answer the question: does it qualitatively describe strongly correlated metals and their superconductivity? 

\begin{figure}[ht!]
\includegraphics[width=\linewidth]{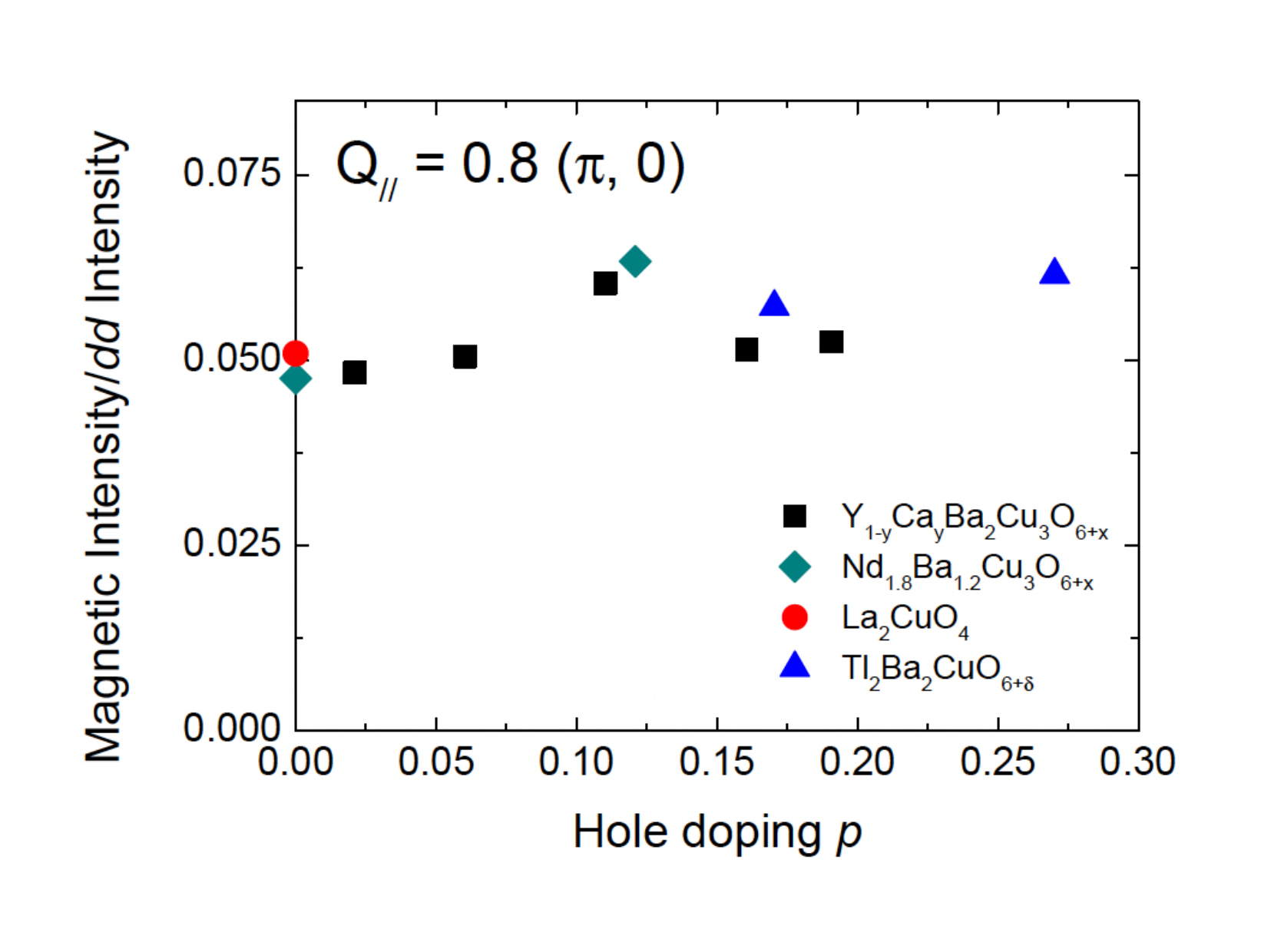}
\caption{Energy-integrated intensity of magnetic excitations (mostly from
RIXS, normalized to the $dd$ excitation intensity) as a function of
doping, for several cuprates. (from Le Tacon et al. ref. \cite{Le2013}).}
\label{fig25}
\end{figure}

 In this light, before mentioning our results for a paradigmatic approach to the $U=\infty$ limit I present a piece of experimental evidence which suggests that at least in hole doped cuprates the electron correlation $U$ continues to be both large and nearly unchanged for hole doping $p$ ranging from $0$ (undoped) to $0.27$ (overdoped), so that this limit small (in (1/U) departures from $U=\infty$ are relevant. The evidence is the (broadened) spin wave spectrum of doped cuprates obtained via inelastic neutron scattering studied using RIXS, the excitation energies being as large as $200$ meV. It is found that for large values of $\textbf{Q}$ the dispersion is unchanged for different values of doping, and is fairly accurately given by the spin wave dispersion from the nearest neighbour Heisenberg model with a $J$ of about $1400K$. This implies strong coupling because the idea of a $J$ meaningful only for large $U$, and also implies that $U$ remains large at short distances since $J(\approx t^2/U)$ is relatively unchanged. An independent confirmation is from the nearly unchanged frequency integrated intensity of the spin wave spectrum plotted for a particular large value of $\textbf{Q}$, namely $\textbf{Q} = 0.8(\pi,0)$, for different cuprates with hole doping ranging from $0$ to $0.27$ (Fig. \ref{fig25}). 

The exploration of the metal for $U=\infty$ has been pioneered by Shastry \cite{Shastry2010} starting from $2010$ and continuing, and investigated most thoroughly him. Shastry (who has named this limit as that of the Extremely Correlated Fermi Liquid or ECFL) uses the field source method of Schwinger, has developed a detailed systematic theoretical structure, and has applied the results to cuprates (with nonzero $J$ where necessary). In recent work, we \cite{Hassan2024} have used an equation of motion method which we describe briefly below. 

The infinitely correlated metal is faithfully described in terms of three quantum states $ | {ia}\rangle$ at each site, namely $|{i0} \rangle$, $| {i\uparrow} \rangle$; $|{i\downarrow} \rangle$ or states with no particle or with one particle having spin up or down respectively. The state with two electrons at a site is infinitely high in energy and need not be included. The matrix elements in this space
of states are the Hubbard operators $X_i^{ab}$, fermionic when $a$
and $b$ differ by an electron, and bosonic when $a$ and $b$ differ
by none. They are not canonical Fermi and Bose operators since
the (anti)commutators are not $c$ numbers, but ($X$) operators.
The Hamiltonian in this space of states is

\begin{eqnarray}
\label{ssm_eq2}
 {H}=\sum_{i,\sigma} (-\mu {X_i^{\sigma \sigma}})+ \sum_{ij,\sigma} {t_{ij}X_i^{\sigma0} X_j^{0\sigma}}.
\end{eqnarray}
This would have been exactly soluble if $X_i^{\sigma 0}$ were a canonical
fermion creation operator ${a^\dagger}_{i\sigma}$. 

\begin{figure}[ht!]
\includegraphics[width=\linewidth]{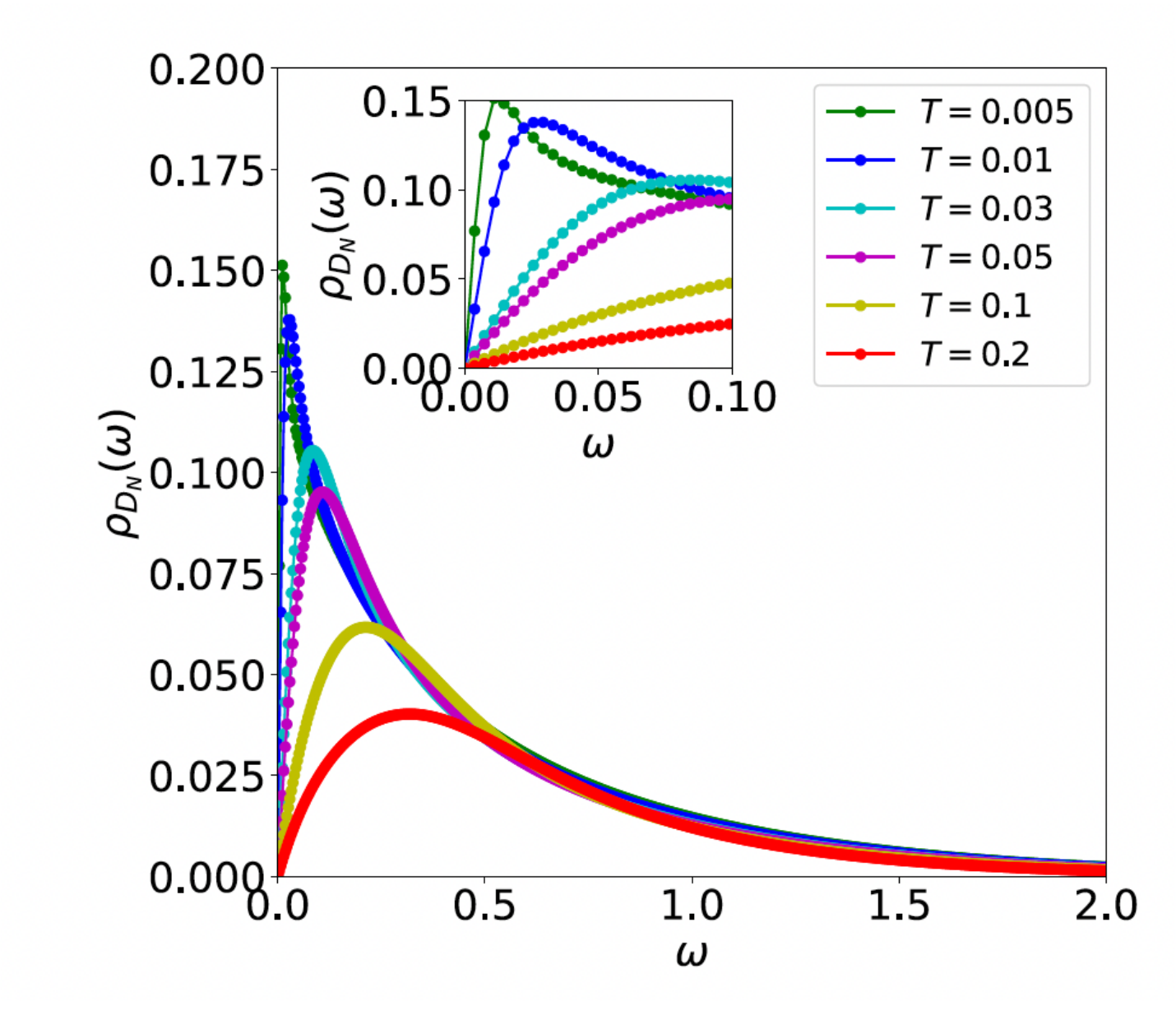}
\caption{Normalized spectral density $\rho_{DN}(\omega)$ of local bosonic fluctuations
for hole doping $x=0.3$ as a function of frequency $\omega$ for various
temperatures $T$ (in units of the nearest neighbour hopping $t$) (from
Hassan at al. ref. \cite{Hassan2024}).}
\label{fig26}
\end{figure}

\begin{figure}[ht!]
\includegraphics[width=\linewidth]{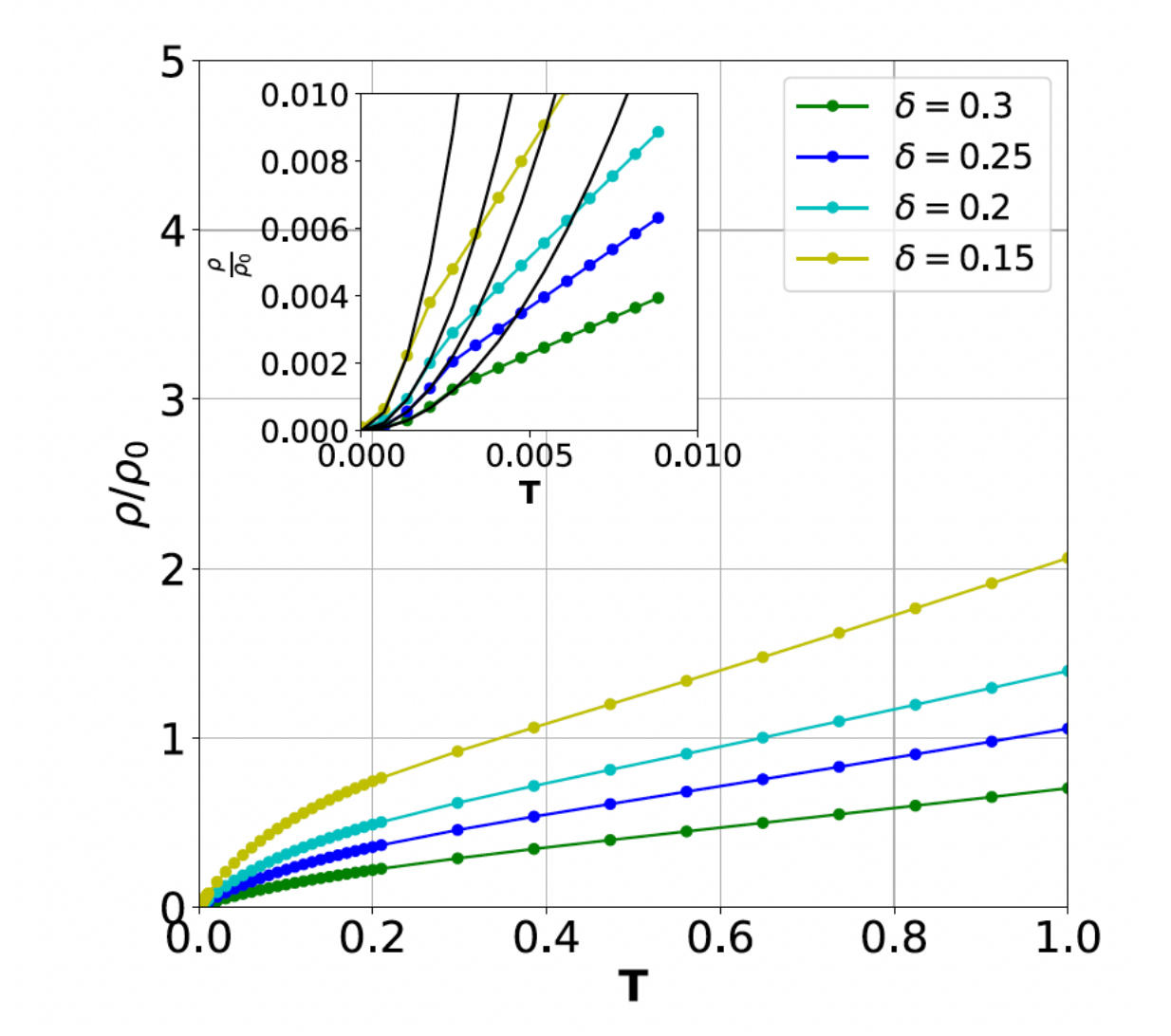}
\caption{Electrical resistivity as a function of temperature, for different values of doping (from Hassan et al., ref. \cite{Hassan2024}).}
\label{fig27}
\end{figure}

In the ECFL, the dynamics of the
electron correlator of the type $\langle {X_i^{\sigma 0}}(t){X_i^{0\sigma}(t^\prime)}\rangle$  involves (e.g. via
the equation of motion) mixed fermionic and bosonic fluctuation
correlation functions which can be decoupled in the $d=\infty$ limit.
The propagation of the resulting bosonic fluctuations (charge
and spin correlators) depend on electron correlators. We
therefore calculate $XX$ correlation functions self consistently.
For example, spectrum of local self generated bosonic
fluctuations obtained this way is shown in Fig. \ref{fig26}.
The diffusive, overdamped form, roughly characterized by a
damping or width but with a long tail and characteristic
broadening with temperature is evident. These fluctuations
are strongly coupled to the electron dynamics, e.g. to the self
energy $\Sigma$. It is seen from exact properties of spectral functions at low energies that $Im \Sigma(\omega,T)$ necessarily has the coherent
liquid form $(\omega^2 + \pi^2 T^2)$ at very low frequencies $\omega$ and
temperatures $T$. This crosses over at very low temperatures
$T\approx 0.003t$ to an incoherent Fermi liquid regime. In measurable
quantities like resistivity, there is a deviation towards linearity
from a ${T^2}$ behaviour starting from this temperature, as seen in Fig. \ref{fig27}.

\begin{figure}[ht!]
\includegraphics[width=\linewidth]{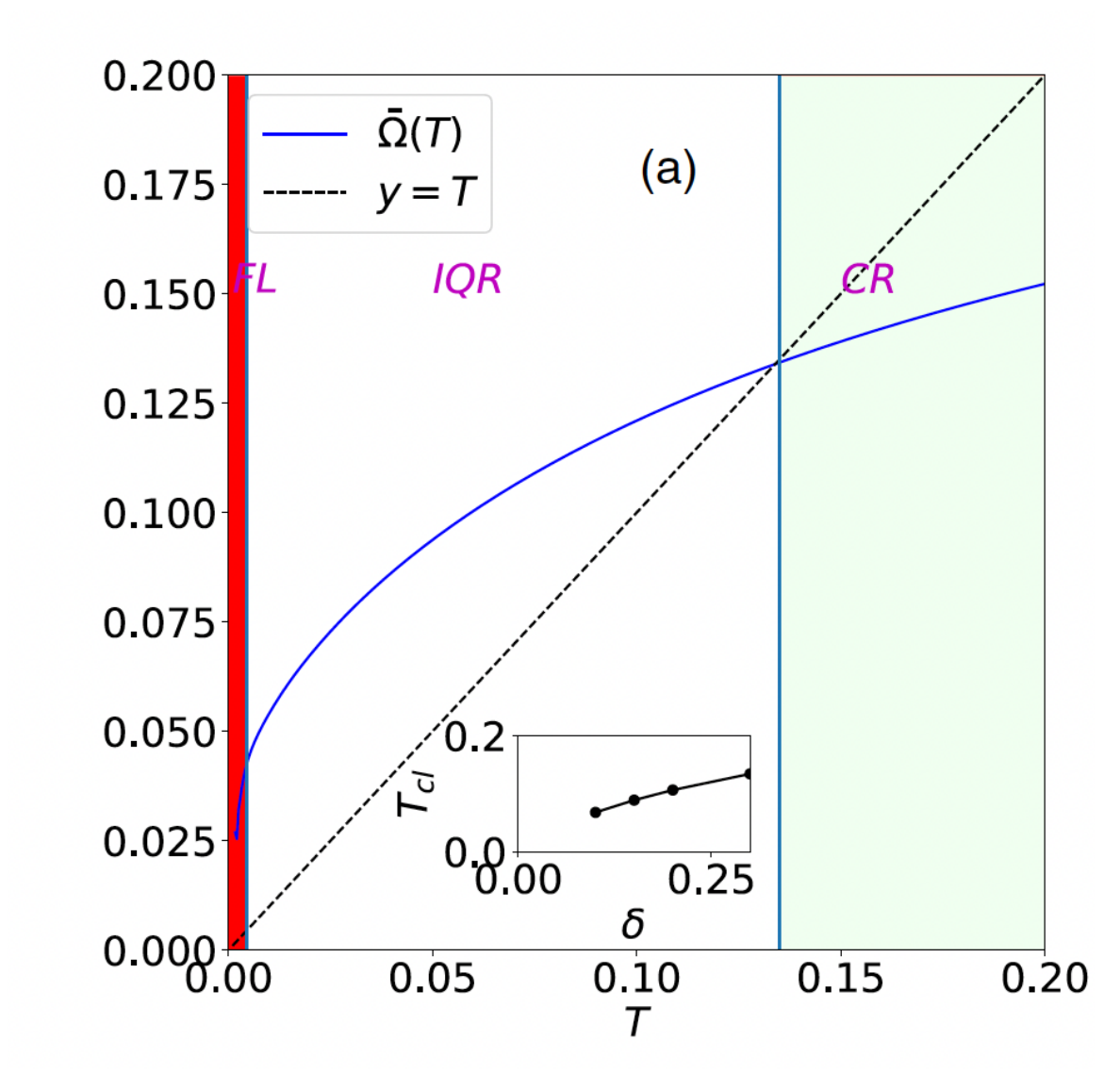}
\caption{Average bosonic fluctuation energy $\Omega(T)=<\omega>$; (in units of $t$) as a
function of temperature (in units of $t$) for doping $x=0.3$. The coherent Fermi liquid $(FL)$, incoherent quantum regime $(IQR)$, and the classical
regime (CR) are also shown (from Hassan et al., ref. \cite{Hassan2024}).}
\label{fig28}
\end{figure}

This figure also shows the ubiquitous linear resistivity
behaviour with the low crossover temperature for it, clearly apparent
in the inset. On a larger temperature scale, the linear behaviour
seems to be composed of two linear segments with a
crossover. Other indications, e.g. from Im $\Sigma(0,T)$ and single
particle spectral density, also support this crossover to an incoherent
Fermi liquid (which has linear resistivity). This seems to
be a paradigmatic behaviour associated with strong correlation.
In terms of the bosonic fluctuations, this is brought
out in Fig. \ref{fig28} where I plot the
average energy $\Omega = \langle \omega \rangle$ of the bosonic fluctuation against
temperature.

One sees clearly the coherent Fermi liquid regime
at very low temperatures. There is a ‘classical’ regime for which $T>\Omega$
. In between, there is a large incoherent Fermi liquid
regime, which is dominated by local quantum bosonic
fluctuations. In this regime the thermal energy is small compared to the
average energy of diffusive fluctuations which can also be
thought of as quantum electrical noise at each lattice site (‘white’ in the
sense that the strength does not depend on the frequency).
We believe that we have unearthed two crucial features of
ECFL: one is that there are strong, local, diffusive, self
generated, bosonic (charge and spin) fluctuations coupled to
electrons and the other is that there is, consequently, a large
incoherent quantum regime. 

However, while it seems plausible
that some unusual properties of the metallic state (e.g. the incoherent Fermi liquid state and ubiquitous linear resistivity) are paradigmatic features of strong correlations
and are captured in a $U=\infty$ theory, we cannot and do not compare its results with real systems yet for several reasons, some of which are the following. The theory is still too opaque, technically weak, and complicated; it needs to be perhaps cast in a different, more accessible form which explicitly brings out our observations and results. We have worked with a paramagnetic state, while the ground state is believed to be a (Nagaoka) ferromagnet at least
for quite low hole densities so that for such densities one is starting with the ‘wrong’ ground state. The crossover from a coherent to an incoherent Fermi liquid occurs at a very low temperature the origin of whose small scale is not clear. Real strongly correlated systems have a large but finite $U$, so that there must be significant $(1/U)$ effects (including $d$-wave superconductivity and many of the phenomena described above). Maybe with a theory which includes $O(1/U)$ effects one can confront experiments. Such a $(1/U)$ perturbation theory is being developed.

To conclude, we have been faced with a strange beast called Cuprate High Temperature Superconductors for many decades (and now, we have many like them). We know
many of its characteristics; some have been described
above. We have been trying (not quite successfully) to
describe them in the language we know. Maybe a different
language is needed.

\section{Acknowledgements}I am very thankful to Professor Tanusri Saha-Dasgupta for the opportunity to present this review like talk at the Bose centenary conference. I am also thankful to Dr. Arijit Haldar and Ms. Monalisa Chatterjee for help with the typescript.

\section*{References} 
\bibliographystyle{iopart-num}
\bibliography{ref}
\end{document}